\documentclass[a4paper,onecolumn,unpublished]{quantumarticle}
\pdfoutput=1
\usepackage{amsmath,amssymb,amsthm,mathtools}
\usepackage{booktabs}
\usepackage{tabularx}
\usepackage{xcolor}
\usepackage{pgfplots}
\pgfplotsset{compat=1.18}
\usepackage[numbers,sort&compress]{natbib}
\usepackage[colorlinks=true,linkcolor=blue,citecolor=blue,urlcolor=blue]{hyperref}

\newtheorem{theorem}{Theorem}
\newtheorem{proposition}[theorem]{Proposition}
\newtheorem{corollary}[theorem]{Corollary}
\newtheorem{lemma}[theorem]{Lemma}
\theoremstyle{remark}

\newcommand{\R}{\mathbb{R}}
\DeclareMathOperator{\artanh}{artanh}
\DeclareMathOperator{\arcosh}{arcosh}

\begin{document}

\title{Degree and normalization frontiers for bounded QSVT inversion}

\author{Hongyu Wu}
\affiliation{Department of Electrical and Computer Engineering, Kansas State University}

\date{17 November 2026}

\maketitle

\begin{abstract}
Quantum singular value transformation (QSVT) implements matrix inversion using bounded polynomials. Increasing polynomial degree can improve normalization and reduce amplified query cost, with savings that depend on the error criterion. However, existing bounded methods either fix the normalization or work at condition numbers far below those used in fault-tolerant resource estimates. We therefore analyze how degree and normalization affect amplified query cost under absolute reciprocal error, which controls inverse-operator accuracy, and under residual error, which suffices for normalized linear solves. For absolute reciprocal error, we establish a scaling limit to a constrained least-norm problem for entire functions and characterize the high-accuracy frontier of this limiting problem using elliptic integrals. Constructions built from near-optimal limiting functions achieve certified query-cost savings of at least 11.8 percent over the rescaled minimax baseline at condition number ten thousand with error one millionth, and at least 22.8 percent at one million with error one ten-billionth. A simpler two-term rule also has certified savings, and its phase factors at condition number ten thousand pass numerical checks at selected points. For residual error, a sparse Chebyshev construction achieves certified savings of at least 8.7 percent over the residual-minimax baseline at condition number ten thousand with normalized-state error at most one hundred-millionth, and at least 13.1 percent at one million with error at most one ten-billionth. These results quantify the savings from jointly choosing degree and normalization relative to the minimax baselines.

\end{abstract}

\section{Introduction}

Quantum algorithms for linear systems, Green's functions and related linear-algebra tasks often require implementing a matrix inverse from a block-encoding. Quantum singular value transformation (QSVT) enables this operation by applying a polynomial approximation of the reciprocal function to the singular values of the encoded matrix \cite{gilyen2019,martyn2021}. However, QSVT requires the polynomial to be bounded in magnitude by one on $[-1,1]$. If $q$ approximates $1/x$ on $[1/\kappa,1]$ for condition number $\kappa$, QSVT implements a scaled polynomial $c\,q$, where the normalization $c$ cannot exceed the reciprocal of the maximum of $|q|$ on $[0,1]$. With amplitude amplification, the query cost depends on both the polynomial degree and the inverse normalization $1/c$. Minimizing degree alone therefore need not minimize the total query cost.

Existing approaches differ in how they treat polynomial boundedness and approximation error. Explicit constructions enforce boundedness by design \cite{childs2017,gilyen2019}. For example, Gily\'en et al.\ multiply the polynomial of Childs, Kothari and Somma by a rectangle function and use normalization $1/(2\kappa)$ \cite{gilyen2019}. In contrast, the minimax polynomial minimizes the uniform approximation error at a fixed degree and is known in closed form \cite{berntson2025exact,sunderhauf2025}. Its approximation properties are related to classical two-interval results, whose asymptotics Eremenko and Yuditskii attribute to Akhiezer \cite{eremenko2011}. At equal error, the fitted degree of the minimax polynomial is smaller than that of the Childs--Kothari--Somma construction by a factor of about $2.6$ \cite{sunderhauf2025}. However, the maximum magnitude of the minimax polynomial inside the spectral gap $(0,1/\kappa)$ exceeds $\kappa$ by a factor that grows slowly with increasing accuracy. The smaller normalization required for QSVT consequently offsets part of the minimax polynomial's degree advantage.

This tension suggests a different strategy. A higher degree can reduce the polynomial's maximum magnitude, which allows a larger normalization and can lower the amplified query cost. Although this trade-off has been recognized, its quantitative behavior remains poorly understood at large condition numbers. S\"underhauf et al.\ discuss the balance between degree, maximum magnitude and windowing cost, but regard an analytic treatment as impractical \cite{sunderhauf2025}. Numerical approaches have also explored bounded polynomial design. Dong, Larsen, Lin and Sarkar use constrained Remez and nonlinear Fourier methods at condition number ten with fixed normalization $1/\kappa$ \cite{dong2026}. Their constrained Remez method checks candidate extrema numerically. Their nonlinear Fourier retraction instead returns phase factors whose polynomial is bounded by one on the full interval in exact arithmetic. This bound does not by itself certify the approximation error of the realized polynomial. Deiml and Peterseim jointly choose the polynomial and its normalization under an $L^2$ error criterion at condition numbers 3 and 5 and degrees up to 16 \cite{deiml2026}. Adhikari formulates an $L^2$ problem with a bound on the polynomial maximum \cite{adhikari2025}. These studies leave open the potential savings from jointly optimizing degree and normalization in the high-condition-number regime.

The error criterion is particularly important in this regime. Fault-tolerant resource estimates consider condition numbers up to $10^6$ with stringent normalized-state accuracy requirements \cite{jennings2023,dalzell2024}. However, a normalized-state error target does not directly specify the required absolute accuracy of the inverse operator. We therefore distinguish two criteria: absolute reciprocal error (criterion A), which controls inverse-operator accuracy, and residual error (criterion B), which suffices for normalized linear solves. Each criterion has its own known minimax baseline \cite{sunderhauf2025}. The saving from a joint choice of degree and normalization is therefore assessed separately under each criterion.

We quantify the savings with a query cost proportional to polynomial degree divided by normalization. Our analysis characterizes the scaling limit under criterion A and establishes certified finite-condition-number savings under both criteria. Each gain we report is measured against the minimax baseline for the same error criterion. Our main contributions are as follows.

\begin{enumerate}
\item We establish a scaling limit for criterion A as the condition number grows with the error held fixed relative to the condition number. In this limit the least feasible polynomial norm, divided by the condition number, converges to the optimum of an extremal problem for entire functions subject to an error constraint outside a fixed gap. We also identify the feasibility threshold. While classical limit theorems connect polynomial approximation errors to entire-function approximation \cite{ganzburg1992limit,ganzburg2023sharpIV}, our result applies to the least norm under a prescribed error constraint.

\item We characterize the high-accuracy frontier of the limiting problem using elliptic integrals and derive the optimal asymptotic cost constant, $2.9492$. The leading term of the limiting gain is $0.1530$ times the square root of the logarithm of the condition-number-to-error ratio.

\item This leading term exceeds one only when that logarithm exceeds $42.7$, so at practical accuracies we directly certify two-sided bounds on the optimal limiting gain at five accuracy levels. The near-optimal entire functions used in our finite constructions achieve limiting gains within 0.1 percent of this optimal gain.

\item We compute finite degree--normalization frontiers under both criteria at condition numbers 20 and 40. We further develop a simpler two-term rule, perform selected-point phase checks, examine windowing costs and evaluate polynomial substitutions in two reproduced cells of a published query model. The near-optimal constructions under criterion A have certified query-cost savings of at least 11.8 percent over the rescaled minimax baseline at condition number $10^4$ with error $10^{-6}$, and at least 22.8 percent at $10^6$ with error $10^{-10}$. The sparse Chebyshev construction under criterion B has certified savings of at least 8.7 percent over the residual-minimax baseline at condition number $10^4$ with normalized-state error at most $10^{-8}$, and at least 13.1 percent at $10^5$ and $10^6$ with error at most $10^{-10}$.
\end{enumerate}

The study map below summarizes the three questions these results address, with the corresponding analyses and evidence.

\begin{center}
\begin{minipage}{\textwidth}
\centering
\textbf{Study map.}\\[4pt]
\small
\begin{tabularx}{\textwidth}{>{\raggedright\arraybackslash}X>{\raggedright\arraybackslash}X>{\raggedright\arraybackslash}X}
\toprule
Question & Analysis and section & Evidence and scope \\
\midrule
How much does joint optimization save? &
Linear programs (III) and finite constructions (V) &
Numerical frontiers and certified savings under criteria A and B \\
\addlinespace
How do savings depend on accuracy and condition number? &
Entire-function limit and asymptotic frontier (IV) &
Distinct limiting regimes and certified bounds at errors held fixed relative to the condition number \\
\addlinespace
What is required for implementation and use? &
Phase factors (V.C), windowing (VI) and cost models (VII) &
Selected-point phase checks, specified window comparisons and conditional application estimates \\
\bottomrule
\end{tabularx}
\end{minipage}
\end{center}

The Supplemental Material provides the proofs, certificates and computational details in Sections S1--S12. The supplement and verification set accompany the arXiv version as ancillary files. Appendix~\ref{app:repro} describes the programs, the verification records and their availability.

\section{Problem setting and cost model}\label{sec:setting}

Let $a=1/\kappa$. We consider odd polynomials in the span of the first $m$ odd Chebyshev terms,
\begin{equation}
q(x)=\sum_{k=0}^{m-1}c_k\,T_{2k+1}(x),
\end{equation}
whose degree is at most $2m-1$. Under criterion A, such a $q$ is feasible for absolute reciprocal error $\varepsilon$ when
\begin{equation}\label{eq:feasible}
|q(x)-1/x|\le\varepsilon\qquad\text{for all }x\in[a,1].
\end{equation}
The error $\varepsilon$ bounds $q-1/x$ itself, not its ratio to $1/x$. For a matrix $A$ with $\|A\|\le1$ and $\|A^{-1}\|\le\kappa$, the inversion algorithm applies $q$ to the singular values of $A$, which lie in $[a,1]$. For $A=U\Sigma V^\dagger$, QSVT applied to a block-encoding of $A^\dagger$ gives $V\,q(\Sigma)\,U^\dagger$, which approximates $A^{-1}=V\Sigma^{-1}U^\dagger$. The computed vector therefore differs from $A^{-1}b$ by at most $\varepsilon\|b\|\le\varepsilon\|A^{-1}b\|$. Hence $\varepsilon$ bounds the relative error of the solution vector. In a solve we take $\varepsilon<1$, so the computed vector is nonzero. Since the circuit applies $c\,q$ and the solution state is normalized, the error of the normalized solution state does not depend on the normalization. This error is at most $2\varepsilon$. The bounded frontier is
\begin{equation}\label{eq:frontier}
M(m,\varepsilon)=\min\Bigl\{\max_{x\in[0,1]}|q(x)| \,:\, q \text{ feasible}\Bigr\}.
\end{equation}
Because $q$ is odd, its maximum magnitude is the same on $[0,1]$ and $[-1,1]$. QSVT can therefore implement $c\,q$ with $c=1/M$.

\paragraph{Cost model.} In a linear solve with amplitude amplification and a unit vector $b$, the QSVT circuit prepares the solution state with an amplitude equal to $c$ times the norm of the computed vector. By the bound above, this norm lies within a factor $1\pm\varepsilon$ of $\|A^{-1}b\|$, which is common to all polynomials. The amplitude is therefore proportional to $c$ up to this factor. The number of amplification rounds is proportional to $1/c$. Each round uses a transformation of degree at most $2m-1$. We count its query budget as $2m-1$, so the amplified query budget is proportional to $(2m-1)/c$. We use
\begin{equation}\label{eq:cost}
C(m)=m\,M(m,\varepsilon)
\end{equation}
as the cost. We assume that the block-encoding of $A$ is exact and count each query to it at the same cost. The certified savings do not cover errors in $A$ or in its block-encoding. State preparation adds the same cost to every amplification round. Including this cost would raise the gains, because the optimized bounded polynomials need fewer rounds than the rescaled minimax baseline. Leaving it out is therefore conservative. The norm of the computed vector can change a ratio of two solve costs by a factor between $(1-\varepsilon)/(1+\varepsilon)$ and its reciprocal. At the certified points of Section~\ref{sec:constructions} this factor differs from one by less than $3\times10^{-6}$. Section~\ref{sec:subroutine} discusses uses of the inverse in which the cost depends on $c$ differently.

\paragraph{Baseline.} The minimax polynomial without the bound is known in closed form \cite{sunderhauf2025}. Sec.~S1 proves the value of its error. With $t(y)=(2y-(1+a^2))/(1-a^2)$, $\beta=(1-a)/(1+a)$ and $\ell_m=T_m+\beta\,T_{m-1}$,
\begin{equation}\label{eq:closedform}
q^*_m(x)=\frac{1}{x}\left[1-\frac{\ell_m(t(x^2))}{\ell_m(t(0))}\right],\qquad
E_m=\max_{x\in[a,1]}|q^*_m(x)-1/x|=\frac{(1-a)^m}{a(1+a)^{m-1}}.
\end{equation}
Let $m_u$ be the least $m$ with $E_m\le\varepsilon$. Let $\rho_u=\max_{[0,1]}|q^*_{m_u}|/\kappa$ be the overshoot of this polynomial in units of $\kappa$. The rescaled minimax polynomial $q^*_{m_u}/(\rho_u\kappa)$ is realizable and costs $m_u\rho_u\kappa$. In the gap $q^*_m$ grows pointwise with $m$ (Corollary~S1.4). At every criterion-A parameter point where this paper reports a gain, ball arithmetic gives $q^*_{m_u}(3a/4)>\kappa+\varepsilon$ (program \texttt{overshoot\_scan.py}). The right-hand side bounds $|q^*_{m_u}|$ over $[a,1]$. The maximum therefore lies in the gap. Every $q^*_m$ with $m>m_u$ then has norm at least $\rho_u\kappa$ and a larger rescaled cost. Hence $m_u$ is also the best degree for rescaling. We report the gain
\begin{equation}\label{eq:gain}
G(\kappa,\varepsilon)=\frac{m_u\,\rho_u\,\kappa}{\min_{m\ge m_u} m\,M(m,\varepsilon)}\ \ge 1 .
\end{equation}
A gain $G$ corresponds to a cost reduction by the fraction $1-1/G$. Using $(2m-1)M$ instead of \eqref{eq:cost} lowers every criterion-A gain reported below by a relative amount of at most $1/(6m_u-2)$ over the degrees we compare, $m\le1.5\,m_u$. This amount is about $10^{-3}$ in Table~\ref{tab:measured}, below $1.5\times10^{-6}$ at $\kappa=10^4$ and below $10^{-7}$ at $\kappa=10^5$ and $10^6$.

\paragraph{Criterion B and its baseline.} Alternatively, require $|xq(x)-1|\le\delta$ on $[a,1]$, with $0<\delta<1$. If $Q=Vq(\Sigma)U^\dagger$, then
\begin{equation*}
Q-A^{-1}=V\bigl(\Sigma q(\Sigma)-I\bigr)V^\dagger A^{-1}.
\end{equation*}
Thus $\|(Q-A^{-1})b\|\le\delta\|A^{-1}b\|$, the normalized-state error is at most $2\delta$ in the 2-norm, and $\|Q-A^{-1}\|\le\kappa\delta$. The cost model above applies with $1\pm\delta$ in place of $1\pm\varepsilon$.

Matrix functions, absolute-error matrix elements and products of inverse block-encodings need the operator guarantee of criterion A \cite[Theorems 17 and 23 and Sec.~3.2]{tong2021}, as does the operator-norm statement of Gily\'en et al.\ \cite[Theorem 41]{gilyen2019}. Normalized solves \cite{childs2017,costa2023,dalzell2024,jennings2023,lefterovici2026} need only criterion B after conversion to the state norm in question. Our $2\delta$ bound uses the 2-norm and does not identify it with a 1-norm or trace-distance target.

The residual-minimax polynomial is \cite[Eq.~(21)]{sunderhauf2025}
\begin{equation*}
q_{R,m}(x)=\frac{1-T_m(t(x^2))/T_m(t(0))}{x},\qquad
\|xq_{R,m}-1\|_{[a,1]}=\operatorname{sech}(2m\operatorname{artanh}a).
\end{equation*}
Its least feasible number of odd terms is $m_R=\lceil\operatorname{arcosh}(1/\delta)/(2\operatorname{artanh}a)\rceil$. Write $\rho_R=\|q_{R,m_R}\|_{[0,1]}/\kappa$. For a feasible degree-$2m-1$ polynomial of norm at most $M$, we report $G_B=m_R\rho_R\kappa/(mM)$. The Chebyshev window of Section~\ref{sec:windowing} is $xq_{R,\ell}(x)$.

The choice of error criterion changes the comparison. At the three criterion-A points of Table~\ref{tab:practical}, equal state guarantees mean $\delta=\varepsilon$. Then the cost of $q_R$ is 0.499 to 0.568 times that of the rescaled criterion-A minimax polynomial and 0.567 to 0.722 times that of our criterion-A near-optimal construction. At equal operator error, $\delta=\varepsilon/\kappa$, the cost of $q_R$ is 1.17 to 1.32 times that of our criterion-A construction. These are numerical evaluations of the stated cost model. We use no other pairing for a comparison of costs across the criteria.

\paragraph{Scaling variables.} For the criterion-A scaling analysis, we write $\mu=m/\kappa$, $\eta=\varepsilon/\kappa$ and $L=\ln(1/\eta)=\ln(\kappa/\varepsilon)$. We call $\eta$ the scaled error because it compares $\varepsilon$ with the largest value $\kappa$ of $1/x$ on $[a,1]$. Because $aE_m\approx e^{-2\mu}$ for small $a$, the least feasible number of odd Chebyshev terms for the unbounded approximation satisfies $m_u/\kappa\approx L/2$. We write $\mu_u=L/2$. At $\kappa=10^4$ with $\varepsilon=10^{-6}$, $L=23.0$ and $m_u/\kappa=11.5$. At $\kappa=10^6$ with $\varepsilon=10^{-10}$, $L=36.8$ and $m_u/\kappa=18.4$. The degree $2m_u-1$ is then $3.7\times10^7$. In the same limit the overshoot of the minimax polynomial tends to
\begin{equation}\label{eq:rhoinf}
\rho_\infty(\mu)=\max_{0<s\le1}h_{2\mu}(s),\qquad
h_\sigma(s)=\frac{1-e^{-\sigma}\bigl[\cosh(\sigma w)+w\sinh(\sigma w)\bigr]}{s},\quad w=\sqrt{1-s^2},
\end{equation}
where $s=x\kappa$ is the scaled gap coordinate. For large $\mu$, $\rho_\infty(\mu)\sim c_0\sqrt{\mu}$ with $c_0=(1-e^{-z_0})/\sqrt{z_0}=0.6381726863$ and $e^{z_0}=1+2z_0$ (Sec.~S4). At finite $\kappa$ we evaluate $\rho_u$ from \eqref{eq:closedform}. It equals $2.69590$ at $(\kappa,\varepsilon)=(10^5,10^{-10})$ and $2.78144$ at $(10^6,10^{-10})$.

\section{Measured trade-off}\label{sec:measured}

\paragraph{Method.} This section measures the gain \eqref{eq:gain} directly, by linear programs at condition numbers up to 40. For given $(m,\varepsilon)$ we solve \eqref{eq:frontier} as a linear program in the coefficients $c_k$ and the bound $M$. The error constraint is imposed on a finite set of points of $[a,1]$. The bound constraint is imposed on a finite set of points of $[0,1]$. Both sets grow by cutting planes taken from a verification grid. The grid uses all Chebyshev nodes of the full degree and, in addition, Chebyshev nodes mapped onto $[a,1]$ and clustered at $x=a$. The clustering is needed because the error of the extremal oscillates near $x=a$ on a scale far below the node spacing of $[-1,1]$. The large local maxima of the error and of $|q|$ on the grid are refined by golden-section search. We use HiGHS 1.15.1 \cite{huangfu2018} with one thread, scaled iterative refinement and a continuation in $\varepsilon$. The reported $M$ is the Ehlich--Zeller upper bound computed from the values at Chebyshev nodes \cite{ehlich1964}. It exceeds the attained maximum by at most $3.1\times10^{-4}$ relative. We compute $\rho_u$ with the same bound. This factor therefore cancels in $G$ to about $10^{-5}$. Accepted solutions satisfy \eqref{eq:feasible} on the verification grid to within $10^{-5}$ relative plus a floor of $4\times10^{-15}\max|1/x|$. At $\eta=10^{-10}$ the floor amounts to $4\times10^{-5}$ relative to the error tolerance. As a control, the program reproduces the exact error $E_m$ of \eqref{eq:closedform} within $4.2\times10^{-6}$ relative in all 9 tested cases, at $\kappa\in\{10,30,100\}$ and $m$ from 10 to 900. Six further controls of the node evaluation and of the Ehlich--Zeller bound also pass. Double precision limits the method to $\eta\ge10^{-10}$. Sec.~S10 gives the details.

\paragraph{Result.} Table~\ref{tab:measured} reports the gain at $\kappa=20$ and $40$. Its column $m_u/\kappa$ is the scaled minimal unbounded number of terms at $\kappa=40$. The column $m^\ast/m_u$ is the ratio of the numbers of odd Chebyshev terms at the minimum cost on the grid. The column $\rho_u\kappa/M$ is the increase of the normalization at that degree. In the table, $G$ and the cost reduction $1-1/G$ are computed from the unrounded gains and rounded down. For each $\eta$ the minimum over $m$ is taken on a grid of twelve values of $m/m_u$ from 1 to 1.5. At $\eta=10^{-10}$ and $\kappa=40$, 9 of the 12 grid points converged. A minimum over a grid can only overestimate the least cost. Up to the stated tolerances, each reported $G$ therefore underestimates \eqref{eq:gain}. The bounded optimum estimated by the LP has a cost 3.4 to 12.1 percent below that of the rescaled minimax baseline. The gain rises steadily with $L$. The gains at the two condition numbers agree within 0.009 in $G$. In four of the six rows $G$ is slightly lower at $\kappa=40$.

The certified point $\kappa=10^4$ with $\varepsilon=10^{-6}$ of Section~\ref{sec:constructions} has $L=23.03$, as in the last row of the table. Measurements at the two condition numbers do not determine the gain at $\kappa=10^4$. As $\kappa$ grows at fixed $\eta$, the gain converges to the limiting gain of Section~\ref{sec:theory}. At this $L$ the limiting gain lies between 1.1382 and 1.1387. Table~\ref{tab:practical} gives certified values at $\kappa=10^4$ for a near-optimal construction and for the two-term rule.

\begin{table}[t]
\caption{LP estimate of the gain of the bounded optimum over the rescaled minimax polynomial under criterion A.}
\label{tab:measured}
\centering
\begin{tabular}{cccccccc}
\toprule
$\eta$ & $L$ & $m_u/\kappa$ & $G$, $\kappa=20$ & $G$, $\kappa=40$ & $m^\ast/m_u$ & $\rho_u\kappa/M$ & cost reduction \\
\midrule
$10^{-7}$ & 16.12 & 8.07 & 1.040 & 1.035 & 1.27 & 1.32 & 3.4--3.9\% \\
$10^{-8}$ & 18.42 & 9.22 & 1.077 & 1.069 & 1.24 & 1.33--1.34 & 6.5--7.2\% \\
$3\times10^{-9}$ & 19.62 & 9.82 & 1.092 & 1.093 & 1.24 & 1.36 & 8.4--8.5\% \\
$10^{-9}$ & 20.72 & 10.38 & 1.101 & 1.099 & 1.24 & 1.36--1.37 & 9.0--9.1\% \\
$3\times10^{-10}$ & 21.93 & 10.97 & 1.126 & 1.117 & 1.21 & 1.35--1.37 & 10.5--11.2\% \\
$10^{-10}$ & 23.03 & 11.53 & 1.137 & 1.137 & 1.21 & 1.38 & 12.0--12.1\% \\
\bottomrule
\end{tabular}
\end{table}

\begin{samepage}
\paragraph{Mechanism.} At the least feasible number of odd terms $m_u$ itself, the LP estimate improves on the overshoot $\rho_u$ by at most 0.14 percent in this range. The larger gains in Table~\ref{tab:measured} therefore require a higher degree. The minimum-cost polynomial on the grid has 21 to 27 percent more odd terms than $m_u$ and in return raises the normalization by 31 to 38 percent. Below $m_u/\kappa=7$ the gain is at most 1.01, because the overshoot $\rho_u$ is itself small there.

\end{samepage}
\paragraph{The sign function behaves differently.} At fixed degree, we obtain a bounded minimax polynomial for the odd sign function by dividing the unbounded minimax polynomial by one plus its uniform error. The sign-function comparison at fixed degree indicates that the large values of $1/x$ near the gap account for the measured reciprocal gains. It does not compare costs across degrees.

\begin{proposition}\label{prop:sign}
For $0<\vartheta<1$, let $\Lambda$ be the least uniform error on $[\vartheta,1]$ of an odd polynomial with $m$ odd terms approximating the constant 1, and assume $\Lambda<1$. Among such polynomials that are also bounded by one in magnitude on $[-1,1]$, the least error on $[\vartheta,1]$ is exactly $2\Lambda/(1+\Lambda)$. It is attained by the unbounded minimax polynomial divided by $1+\Lambda$.
\end{proposition}

The proof is in Appendix~\ref{app:sign}. Linear programs at $\vartheta\in\{0.01,0.02,0.05,0.1\}$ agree with $2\Lambda/(1+\Lambda)$ within $2.1\times10^{-5}$ relative in 46 of 48 cases. The other two lie at the double-precision floor (Sec.~S10). Imposing the magnitude bound of one costs about $\ln2/(2\artanh\vartheta)$ additional odd terms. At $\vartheta=0.01$ this is 17 percent of the degree at error $10^{-2}$ and 4 percent at $10^{-8}$ in our linear programs. We did not find this identity in the sources we searched. It may nevertheless be known. For the reciprocal, the analogous fixed-degree rescaling reduction does not hold because the unbounded extremal overshoots inside the gap to $\rho_u\kappa$, far above its values on $[a,1]$.

\paragraph{Residual-error frontier.} Table~\ref{tab:residual_lp} gives the corresponding results under criterion B. The LP uses $t(x^2)$ as its polynomial variable, so it constrains the residual directly. A formulation in $x$ loses accuracy when coefficients of order $\kappa$ cancel to a residual of order $\delta$, and fails at $\delta\le10^{-9}$. The tabulated results are LP estimates of the bounded optimum. They are not interval certificates. The gains use the Ehlich--Zeller norm bound. The residual Ehlich--Zeller upper bounds of the LP solutions exceed $\delta$ by up to $0.031$ percent. At $\delta=10^{-6}$ the estimated gains are 1.0042 and 1.0083, and at $10^{-5}$ and $10^{-4}$ the estimates agree with the baseline within 0.0020. With $\delta=\eta$ from $10^{-7}$ downwards, the B gains exceed those of Table~\ref{tab:measured} by 0.005 to 0.020. This pairing equates the tolerances at the gap edge $s=1$. In scaled coordinates the constraints are $|f(s)-1/s|\le\eta$ under A and $|f(s)-1/s|\le\delta/s$ under B for $s\ge1$. Sec.~S12 gives the method and independent checks.

\begin{table}[t]
\caption{LP estimates under criterion B.}
\label{tab:residual_lp}
\centering
\begin{tabular}{llll}
\toprule
$\delta$ & $G_B$, $\kappa=20$ & $G_B$, $\kappa=40$ & $m^\ast/m_R$ \\
\midrule
$10^{-7}$ & 1.0467 & 1.0511 & 1.27 \\
$10^{-8}$ & 1.0842 & 1.0893 & 1.24 \\
$3\times10^{-9}$ & 1.1071 & 1.1050 & 1.23 \\
$10^{-9}$ & 1.1155 & 1.1191 & 1.22 \\
$3\times10^{-10}$ & 1.1352 & 1.1333 & 1.21 \\
$10^{-10}$ & 1.1425 & 1.1440 & 1.20 \\
\bottomrule
\end{tabular}
\end{table}

\section{Scaling limit and the asymptotic frontier}\label{sec:theory}

\subsection{The limiting problem}

The linear programs of Section~\ref{sec:measured} stop at $\kappa=40$, far below the condition numbers of resource estimates. We therefore study criterion A as $\kappa\to\infty$ at fixed scaled error. In this limit $\varepsilon=\eta\kappa$ tends to infinity, so this is an approximation-theoretic scaling limit and not a family of increasingly accurate solves. For $\mu>0$ and $0<\eta<1$, let
\begin{equation}\label{eq:Phi}
\Phi(\mu,\eta)=\inf\Bigl\{\|f\|_{\R} \,:\, f \text{ odd entire of exponential type at most } 2\mu,\ |f(s)-1/s|\le\eta \text{ for all } s\ge1\Bigr\}.
\end{equation}
Here $\|f\|_{\R}=\sup_{s\in\R}|f(s)|$. The substitution $F(s)=a\,q(as)$ maps a feasible polynomial with $m$ odd terms to an odd function that approximates $1/s$ on $[1,\kappa]$ with error $\eta$. The Bernstein--Walsh growth of $F$ tends to exponential type $2\mu$ as $a\to0$.

\begin{theorem}[Scaling limit]\label{thm:limit}
Fix $0<\eta<1$, and let $a_j\to0$ and $m_ja_j\to\mu$. If $\mu>L/2$, then $a_jM(m_j,\eta/a_j)\to\Phi(\mu,\eta)$. If $\mu=L/2$, the same convergence holds along every sequence of feasible finite problems, that is, with $(1+a_j)\beta_j^{m_j}\le\eta$ for $\beta_j=(1-a_j)/(1+a_j)$. However, along some sequences with $m_ja_j\to L/2$, the finite problems are eventually all infeasible. If $\mu<L/2$, the finite problems are eventually infeasible. For $\mu\ge L/2$ the infimum in \eqref{eq:Phi} is attained.
\end{theorem}
The threshold case uses Proposition~S2.4, which proves with the dual identities \eqref{eq:dual} that $h_L$ is the unique feasible function at $\mu=L/2$.

Sec.~S2 proves Theorem~\ref{thm:limit}. Fix constants $\gamma_0>0$ and $C_0>0$. For margins $2\mu-L\ge\gamma_0$ and types $\gamma_0\le2\mu\le C_0\ln(1/a)$, explicit transfer bounds give
\begin{equation}
|aM-\Phi|=O\bigl(a^{2/3}[\ln(1/a)]^{5/2}\bigr),
\end{equation}
where the implied constant depends on $\gamma_0$ and $C_0$. Theorem~\ref{thm:limit} is a limit theorem of the kind studied by Bernstein and Ganzburg, which relates polynomial approximation on dilated intervals to approximation by entire functions of exponential type \cite{ganzburg1992limit,ganzburg2023sharpIV}. For $|x|^\alpha$, Lubinsky proved that the rescaled best polynomials converge to the unique best entire function \cite{lubinsky2007bernstein}. The results of this kind that we read treat the best-approximation error over the whole dilated interval for a target that is locally bounded on the real line \cite[Sec.~7]{ganzburg2023sharpIV}. Theorem~\ref{thm:limit} instead treats a least norm under an error constraint for the target $1/s$, whose pole lies at the end of the gap $[0,1)$ that remains fixed under the scaling. The limiting least norm $\Phi$ is also infinite below the threshold $\mu=L/2$ and finite at that threshold. We did not find a limit theorem of this kind with a least norm under a constraint, a pole at the end of the gap, or such a threshold.
The limiting class also has an exact dual description. Let $\sigma=2\mu$. Take the nodes $s_0=1$ and $s_j=\sqrt{1+v_j^2}$, where $\sigma v_j+\arctan v_j=j\pi$, and the weights $w_0=1/(\sigma+1)$ and $w_j=2/(\sigma s_j^2+1)$. Then every bounded odd entire $f$ of type at most $\sigma$ satisfies
\begin{equation}\label{eq:dual}
\sum_{j\ge0}w_j=1,\qquad \sum_{j\ge0}(-1)^jw_j\,f(s_j)=0,\qquad \sum_{j\ge0}\frac{(-1)^jw_j}{s_j}=e^{-\sigma}.
\end{equation}
The signed measure with mass $(-1)^jw_j$ at each $s_j$ therefore certifies that no such $f$ approximates $1/s$ on $[1,\infty)$ with error below $e^{-\sigma}$. The limit $h_\sigma$ of the minimax polynomials attains that error. The identities \eqref{eq:dual} also yield a convergent hierarchy of finite linear programs for $\Phi$. The certified lower bounds from this hierarchy and the explicit constructions of Section~\ref{sec:constructions} and Sec.~S7.1 enclose $\Phi$ at five bandwidths in intervals of relative width below $0.038$ percent. Sec.~S3 proves the identities and the convergence of the hierarchy. Sec.~S7 lists the intervals.

\subsection{The asymptotic frontier}

Let $D(\eta)=\min_{\mu\ge\mu_u}(\mu/\mu_u)\,\Phi(\mu,\eta)$ be the optimized limiting cost relative to $\mu_u=L/2$. The limiting gain is then $G_\infty(\eta)=\rho_\infty(L/2)/D(\eta)$. The behaviour of $\Phi$ along a fixed degree ratio $\lambda=\mu/\mu_u$ is governed by the action
\begin{equation}\label{eq:action}
R(b)=\int_b^1\sqrt{\frac{x^2-b^2}{1-x^2}}\,dx=E(1-b^2)-b^2K(1-b^2),\qquad 0<b<1,
\end{equation}
where $K$ and $E$ are the complete elliptic integrals of the first and second kind, with the parameter as argument. Since $R$ decreases strictly from 1 to 0, $R(b_\lambda)=1/\lambda$ has a unique root $b_\lambda$ for every $\lambda>1$.

\begin{theorem}[Asymptotic frontier]\label{thm:frontier}
For every fixed $\lambda>1$,
\begin{equation}
\Psi(\lambda)=\lim_{L\to\infty}\Phi\bigl(\lambda L/2,\,e^{-L}\bigr)=\frac{1}{b_\lambda},
\qquad
\frac{\lambda}{\sqrt{\lambda^2-1}}<\Psi(\lambda)<\frac{\lambda}{\lambda-1}.
\end{equation}
\end{theorem}

Sec.~S5 proves Theorem~\ref{thm:frontier}, with both strict inequalities, and Corollary~\ref{cor:constants} below. The lower bound $\liminf\Phi\ge1/b_\lambda$ follows from an inequality that holds before any limit.

\begin{proposition}[Finite lower bound]\label{prop:lower}
Let $f$ be odd entire of type at most $\sigma$ with $\|f\|_\R=B$ and $|f(s)-1/s|\le e^{-L}$ for $s\ge1$. For $0<b<s<1$ with $L\ge\sigma R(b)$, put $J_b(s)=\int_b^s\sqrt{(x^2-b^2)/(1-x^2)}\,dx$ and $T=\exp[-\sigma J_b(s)+\arcosh(s/b)]$. Then
\begin{equation}\label{eq:lower}
B\ge\frac{1-T}{s+bT}.
\end{equation}
\end{proposition}

Appendix~\ref{app:lower} proves the bound with a harmonic majorant on the plane cut along $|x|\ge1$ and $|x|\le b$. With $\sigma=\lambda L$ and $b=b_\lambda$, the factor $T$ tends to zero for every fixed $s>b_\lambda$. Hence $\liminf\Phi\ge1/s$. Letting $s\downarrow b_\lambda$ gives the lower bound. The bound also shows that, along any family of feasible functions with bounded norm, $f(s)\to1/s$ on $(b_\lambda,1)$ as $L$ grows. The norm is set by continuing the reciprocal down to $b_\lambda$.

The matching upper bound uses a weighted sum of squared prolate spheroidal functions \cite{slepian1961} as a kernel $H$, with $f=(1-H)/s$. The local density of $H$ follows from smooth semiclassical functional calculus on the sphere \cite[arXiv v2, Theorem~4.5]{kuster2015}. The exterior decay of $H$ follows from an ordinary differential equation comparison. Appendix~\ref{app:upper} outlines the construction and states the cited theorem in the form used. This construction gives no quantitative threshold in $L$ for the accuracy of the asymptotic upper bound.

Harmonic majorants are a standard tool for extremal problems for entire functions \cite{eremenko2012comb}. Closely related two-interval constructions with elliptic periods appear in the best approximation of the sign function \cite{eremenko2011}. The configuration of two rays and a central segment and the resulting action~\eqref{eq:action} are specific to the present problem. We did not find the frontier of Theorem~\ref{thm:frontier}, or the extremal problem \eqref{eq:Phi}, in the sources on approximation by entire functions and on constrained polynomial approximation that we searched. The closest problem we found, described in \cite{eremenko2012comb}, imposes different bounds on the two half-lines on functions of exponential type one. It is a different extremal problem.

\begin{corollary}[Optimized constants]\label{cor:constants}
The asymptotic cost constant is $\lim_{\eta\to0}D(\eta)=D_\ast=\min_{\lambda>1}\lambda\Psi(\lambda)=2.949189861$, attained at the unique $b_\ast=0.5465094138$ with $E(1-b_\ast^2)=2b_\ast^2K(1-b_\ast^2)$, that is $\lambda_\ast=1/R(b_\ast)=1.611760022$ and $\Psi(\lambda_\ast)=1/b_\ast=1.829794647$. Consequently
\begin{equation}
\lim_{\eta\to0}\frac{G_\infty(\eta)}{\sqrt{L}}=\frac{c_0}{\sqrt2\,D_\ast}=0.1530102351.
\end{equation}
\end{corollary}

Theorem~\ref{thm:frontier} holds for each fixed $\lambda$. To obtain the limit of $D(\eta)$, we also need to control the minimum over $\lambda$. Sec.~S5 shows that the convergence is uniform on compact subintervals of $(1,\infty)$. It also shows that near-minimizing $\lambda$ stay in such a subinterval, which gives $D(\eta)\to D_\ast$. In finite terms the asymptotically optimal design uses about $0.806\,\kappa L$ odd terms and has $M\approx1.830\,\kappa$. The rescaled minimax polynomial uses $\kappa L/2$ odd terms and has $M\approx c_0\kappa\sqrt{L/2}$. The two curves of Theorem~\ref{thm:frontier} alone give the brackets $[2,4]$ for $D_\ast$ and $[0.1128,0.2257]$ for the coefficient (Sec.~S5). The exact values lie inside them.

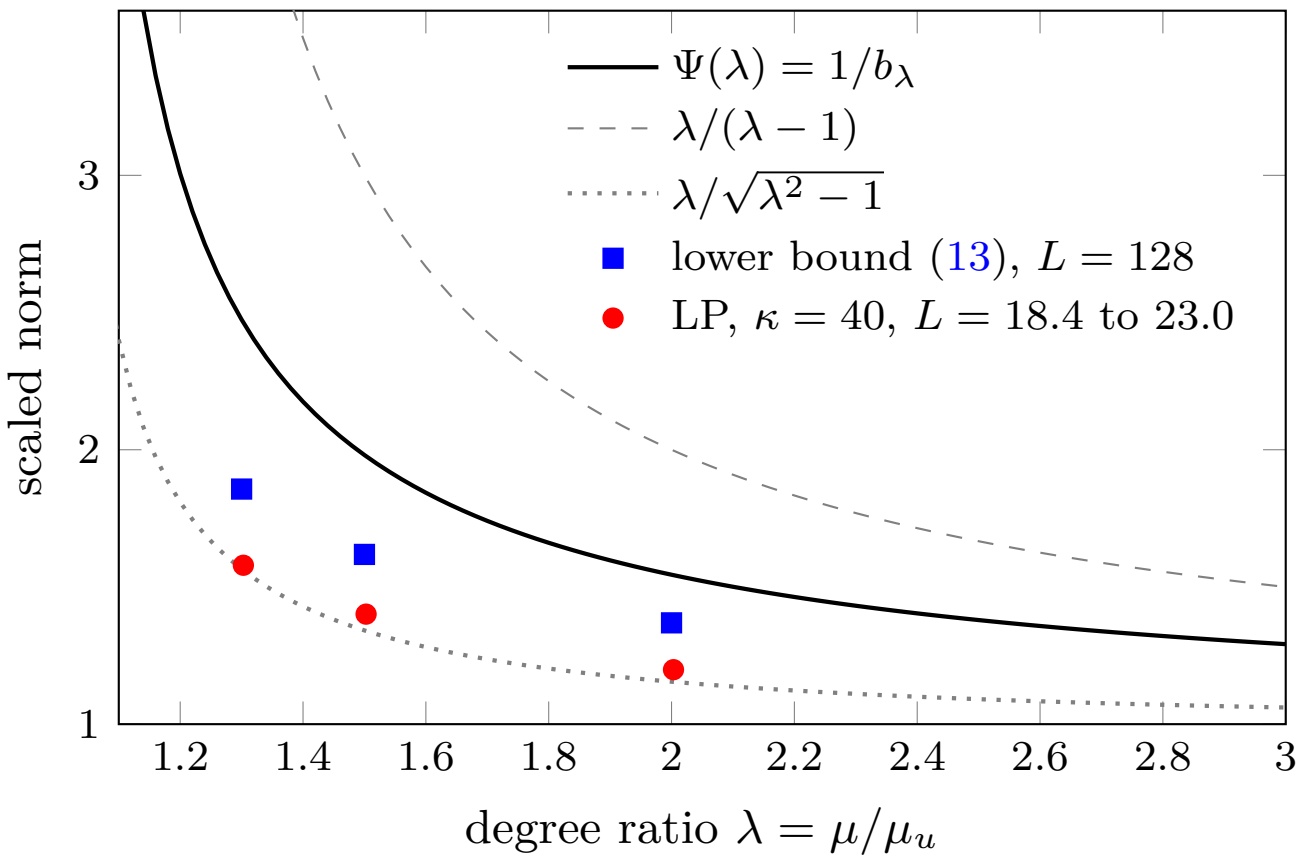
\begin{figure}[t]
\centering
\begin{tikzpicture}
\begin{axis}[width=0.62\textwidth,height=0.42\textwidth,xmin=1.1,xmax=3,ymin=1,ymax=3.6,
  xlabel={degree ratio $\lambda=\mu/\mu_u$},ylabel={scaled norm},legend cell align=left,
  legend style={font=\footnotesize,at={(0.98,0.98)},anchor=north east,draw=none,fill=none},
  tick label style={font=\footnotesize},label style={font=\small}]
\addplot[thick,black] table[x=lam,y=Psi] {data/fig_frontier_psi.dat};
\addlegendentry{$\Psi(\lambda)=1/b_\lambda$}
\addplot[dashed,gray] table[x=lam,y=upper] {data/fig_frontier_psi.dat};
\addlegendentry{$\lambda/(\lambda-1)$}
\addplot[dotted,thick,gray] table[x=lam,y=lower] {data/fig_frontier_psi.dat};
\addlegendentry{$\lambda/\sqrt{\lambda^2-1}$}
\addplot[only marks,mark=square*,mark size=1.8pt,blue] table[x=lam,y=Phi] {data/fig_frontier_l128.dat};
\addlegendentry{lower bound \eqref{eq:lower}, $L=128$}
\addplot[only marks,mark=*,mark size=1.8pt,red] table[x=lam,y=Phi] {data/fig_frontier_lp40.dat};
\addlegendentry{LP, $\kappa=40$, $L=18.4$ to $23.0$}
\end{axis}
\end{tikzpicture}
\caption{Asymptotic frontier of the scaled norm $\Phi(\lambda L/2,e^{-L})$ as $L\to\infty$ (solid), with the two bounding curves of Theorem~\ref{thm:frontier}. Squares are rigorous lower bounds from Proposition~\ref{prop:lower} at $L=128$. Circles are linear-program values $M/\kappa$ at $\kappa=40$, at $L=23.0$ and $18.4$. The computed values lie below the limit and approach it slowly.}
\label{fig:frontier}
\end{figure}

\subsection{What the limit says at practical accuracies}\label{sec:practicallimit}

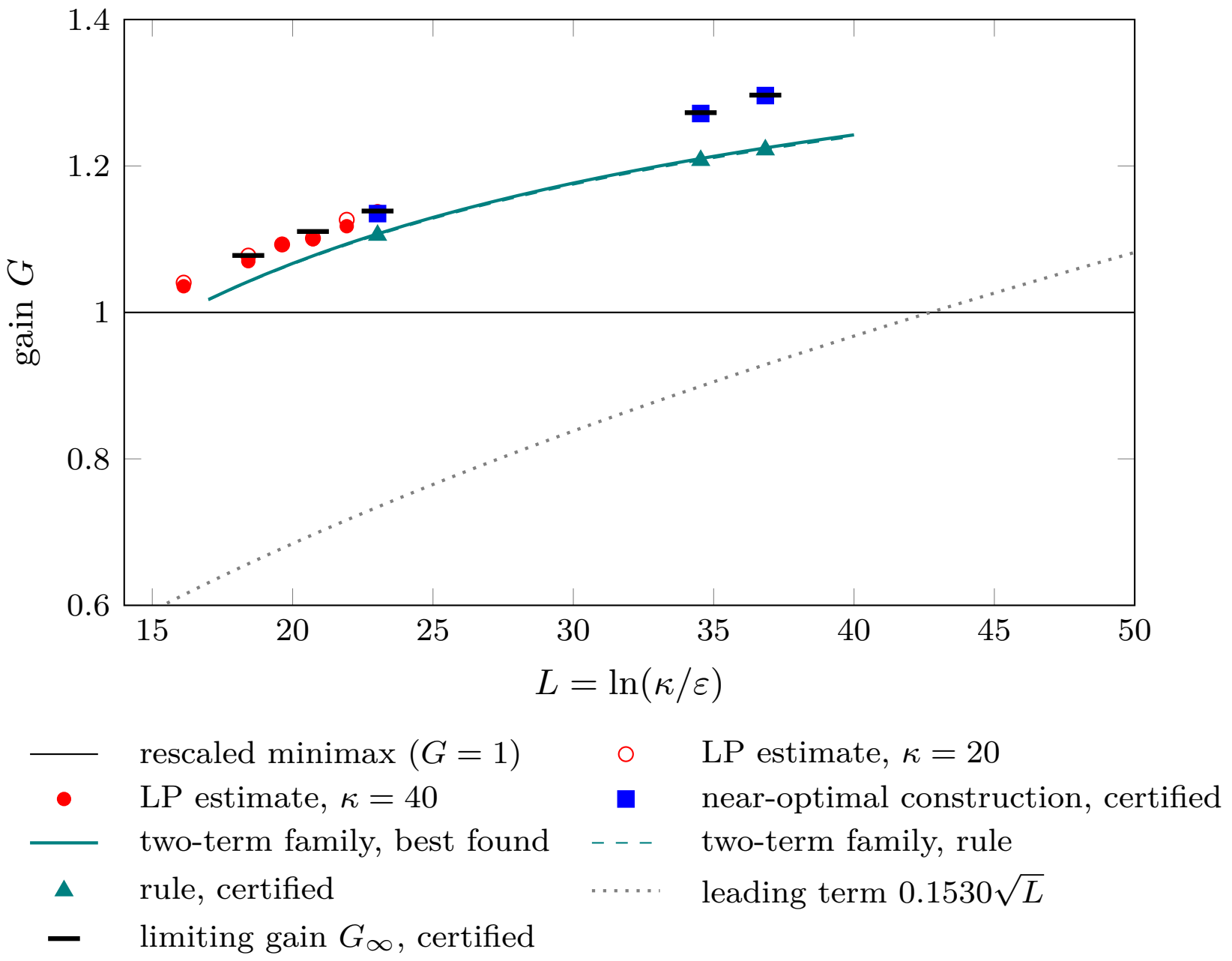
\begin{figure}[t]
\centering
\begin{tikzpicture}
\begin{axis}[width=0.7\textwidth,height=0.45\textwidth,xmin=14,xmax=50,ymin=0.6,ymax=1.4,
  xlabel={$L=\ln(\kappa/\varepsilon)$},ylabel={gain $G$},
  legend columns=2,legend cell align=left,
  legend style={font=\footnotesize,at={(0.5,-0.2)},anchor=north,draw=none,fill=none,column sep=8pt},
  tick label style={font=\footnotesize},label style={font=\small}]
\addplot[thin,black] coordinates {(14,1) (50,1)};
\addlegendentry{rescaled minimax ($G=1$)}
\addplot[only marks,mark=o,mark size=1.8pt,red] table[x=L,y=G] {data/fig_gain_lp20.dat};
\addlegendentry{LP estimate, $\kappa=20$}
\addplot[only marks,mark=*,mark size=1.6pt,red] table[x=L,y=G] {data/fig_gain_lp40.dat};
\addlegendentry{LP estimate, $\kappa=40$}
\addplot[only marks,mark=square*,mark size=2pt,blue] table[x=L,y=G] {data/fig_gain_strong.dat};
\addlegendentry{near-optimal construction, certified}
\addplot[thick,teal] table[x=L,y=G] {data/fig_gain_family.dat};
\addlegendentry{two-term family, best found}
\addplot[dashed,teal] table[x=L,y=G] {data/fig_gain_rule.dat};
\addlegendentry{two-term family, rule}
\addplot[only marks,mark=triangle*,mark size=2.4pt,teal] table[x=L,y=G] {data/fig_gain_rulecert.dat};
\addlegendentry{rule, certified}
\addplot[dotted,thick,gray] table[x=L,y=G] {data/fig_gain_asym.dat};
\addlegendentry{leading term $0.1530\sqrt{L}$}
\addplot[only marks,mark=-,mark size=4pt,black,mark options={line width=1.2pt}] table[x=L,y=mid] {data/fig_gain_ginf.dat};
\addlegendentry{limiting gain $G_\infty$, certified}
\end{axis}
\end{tikzpicture}
\caption{Gain over the rescaled minimax polynomial as a function of $L=\ln(\kappa/\varepsilon)$. LP estimates of the bounded optima at $\kappa=20$ and $40$ (circles), certified near-optimal constructions at $\kappa=10^4$ with $\varepsilon=10^{-6}$ and at $\kappa=10^5$ and $10^6$ with $\varepsilon=10^{-10}$ (squares), the limiting gain $G_\infty$ as $\kappa\to\infty$ at fixed $L$ (black dashes, which are thicker than the certified brackets), the two-term family at $\kappa=10^5$ (solid, best found, and dashed, rule \eqref{eq:rule}) with the certified rule points at $\kappa=10^4$, $10^5$ and $10^6$ (triangles), and the leading term of Corollary~\ref{cor:constants} (dotted). The rule curve lies within 0.22 percent of the best values found and is nearly indistinguishable from the solid curve. The leading term lies below every finite curve in the practical range. The limiting gain itself is at least one there.}
\label{fig:gain}
\end{figure}

Two limits appear in this paper. The first is $\kappa\to\infty$ at fixed $\eta$ (Theorem~\ref{thm:limit}). It gives the limiting gain $G_\infty(\eta)=\rho_\infty(L/2)/D(\eta)$. Sec.~S2 shows that $G(\kappa,\eta\kappa)\to G_\infty(\eta)$ for $L\ge8$. Unlike $G$ in \eqref{eq:gain}, $G_\infty$ is not at least one by definition, because $\Phi$ is a norm on all of $\R$ whereas $\rho_\infty$ is a maximum over the gap. It is at least one for $L\ge8$, which covers the criterion-A Tables~\ref{tab:measured}, \ref{tab:practical} and~\ref{tab:rule} and Fig.~\ref{fig:gain}. The function $h_L$ is feasible at $\mu=L/2$. For $L\ge8$ its norm on $\R$ equals $\rho_\infty(L/2)$ (Sec.~S2), so $D(\eta)\le\Phi(L/2,\eta)\le\rho_\infty(L/2)$.

Explicit entire functions of the type used in Section~\ref{sec:constructions} bound $G_\infty$ from below (Sec.~S7). Because $\Phi$ does not increase with the type, lower bounds on $\Phi$ at finitely many degree ratios give an upper bound on $G_\infty$ (Sec.~S3.6). Sec.~S3 derives from the dual identities \eqref{eq:dual} the explicit lower bound $\Phi(\sigma/2,\eta)\ge\sup_{0<s\le1}\bigl[h_\sigma(s)-(\eta-e^{-\sigma})\,s\cosh(\sigma\sqrt{1-s^2})\bigr]$. We evaluate it on a mesh of degree ratios. Near the minimizing degree ratio, this explicit lower bound is loose, so we add certified lower bounds from the finite linear programs of the dual hierarchy. The certified brackets are $[1.0775,1.0781]$ at $L=18.4$, $[1.1102,1.1108]$ at $L=20.7$, $[1.1382,1.1387]$ at $L=23.0$, $[1.2723,1.2733]$ at $L=34.5$ and $[1.2962,1.2973]$ at $L=36.8$ (Fig.~\ref{fig:gain}). The upper ends bound the optimal limiting gain over all degree ratios. The three certified points of Section~\ref{sec:constructions} have $L=23.0$, $34.5$ and $36.8$. At these values, no bounded polynomial cuts the cost of the rescaled minimax polynomial by more than 12.2, 21.5 and 23.0 percent in this limit. In this limit, the entire functions of that section reduce this cost by at least 12.1, 21.4 and 22.8 percent. Each upper end exceeds its lower end by less than 0.1 percent. The limiting gains of these entire functions are therefore within 0.1 percent of $G_\infty$, the optimum over all degree ratios.

The second limit is $L\to\infty$ (Theorem~\ref{thm:frontier} and Corollary~\ref{cor:constants}). This approximation is poor at the practical accuracies considered here. Its leading term $0.1530\sqrt{L}$ exceeds one only for $L>42.7$, that is for $\eta<2.9\times10^{-19}$. At $L=34.5$ and $36.8$ the leading term is $0.899$ and $0.929$. The certified gains of Section~\ref{sec:constructions} there are $1.2716$ and $1.2961$. The computed values in Fig.~\ref{fig:frontier} lie below the limit and approach it slowly. The circles lie at integer degrees, at $\lambda=1.303$ and $1.503$ for $L=23.0$ and at $\lambda=2.003$ for $L=18.4$. At $\lambda=1.3$ the rigorous lower bound gives $1.856$ at $L=128$, and the limit is $2.474$. The linear program gives $1.579$ at $L=23.0$ and $\lambda=1.303$. The difference between the lower bound \eqref{eq:lower} and the asymptotic frontier decays only as $O((\ln\sigma/\sigma)^{2/3})$. The limit theory therefore determines the scaling and the constants. Practical values require finite computation.

\begingroup
\widowpenalty=\csname @M\endcsname\relax
\paragraph{Finite condition number.} At fixed interior $(\mu,\eta)$, a negative correction to $aM-\Phi$ is at most of order $a^2$. This bound holds along $a=\mu/m$ with integer $m$, where $ma=\mu$ exactly. A positive correction of order $a$ occurs at exact $ma=\mu$ in the following example. At $\mu=1/2$ and $\eta=9/10$ the problem is solved exactly by a signed Chebyshev polynomial, with $\Phi=1/(10\sin1)$ and a positive first-order correction $\cos1/(10\sin^21)$ per unit of $1/\kappa$. Our linear program reproduces this correction within $9\times10^{-10}$ relative at $m$ from 2 to 32. Rounding $m$ so that $ma=\mu+O(a)$ can add a further term of order $a$. Sec.~S6 gives the proofs. Neither the sign nor the leading coefficient of the correction at small $\eta$ is established. The brackets of $G_\infty$ above therefore apply to the limiting gain, not to the gain at the three points of Section~\ref{sec:constructions}. That section certifies lower bounds on the gain at these points.

\par\endgroup
\paragraph{Validation.} We recomputed every constant of Theorem~\ref{thm:frontier} and Corollary~\ref{cor:constants} independently in 50-digit arithmetic. We reproduced the rigorous table of Proposition~\ref{prop:lower} at $L=128$ and $256$, which Sec.~S5 prints with the certified intervals of the constants. We also checked that \eqref{eq:lower} lies below all 19 linear-program values at $\kappa=40$ and below the five certified intervals of the dual hierarchy.

\paragraph{The residual-error limit remains open.} The theory of this section applies to criterion A. The corresponding problem under criterion B is to minimize $\|f\|_{\mathbb R}$ over odd entire functions of exponential type at most $\sigma$, bounded on $\mathbb R$, with $|sf(s)-1|\le\delta$ for $|s|\ge1$. The scaling limit of $q_R$ is
\begin{equation*}
f_{R,\sigma}(s)=\frac{1-\cosh(\sigma\sqrt{1-s^2})/\cosh\sigma}{s}.
\end{equation*}
It is feasible exactly when $\cosh\sigma\ge1/\delta$. We do not prove convergence of the B frontier or an asymptotic cost constant. The B results of Sections~\ref{sec:measured} and~\ref{sec:residual_construction} are finite LP estimates and certificates.

\section{Polynomials at practical condition numbers}\label{sec:constructions}

For criterion A at $\kappa=10^4$ to $10^6$ the degree is $2.8\times10^5$ to $4.7\times10^7$, beyond any direct linear program. We give two kinds of explicit polynomial and certify both.

\subsection{Near-optimal constructions}

An odd entire function of type $\sigma$ is built from 321 coefficients,
\begin{equation}
F(s)=\frac{1-\eta\bigl[\sum_{j=0}^{160}a_j\cos(j\Delta v)+v\sum_{j=1}^{160}b_j\sin(j\Delta v)\bigr]}{s},\qquad v^2=s^2-1,\quad \Delta=\sigma/160,
\end{equation}
with $a_{160}$ fixed analytically so that $F$ is entire. The coefficients were optimized numerically. The bandwidth was found by a numerical search. The optimized function has a certified exterior error of at most $1.0004\,\eta$. Mixing in $h_\sigma$ with weight about $4\times10^{-4}$ gives a function $f$ with exterior error at most $\eta$. The norm of $f$ lies within $0.038$ percent of $\Phi$ at its bandwidth. Near-optimal refers to this property at the chosen bandwidth. The gain of $f$ is less than 0.1 percent below the limiting gain $G_\infty$, which is the optimum over all bandwidths (Section~\ref{sec:practicallimit}). A second mixture with weight about $10^{-6}$ gives $\bar f$ with exterior error at most $(1-10^{-6})\eta$, which reserves $10^{-6}\eta$ for the truncation. The polynomial is $a^{-1}$ times the truncation at odd degree $2m-1$ of the exact Chebyshev expansion of $\bar f(x/a)$. The Bernstein growth bound on an ellipse keeps the truncation error below the reserve. Table~\ref{tab:practical} lists the result.

We certified the exterior error and the norm with our own code built on the Arb library \cite{johansson2017}. The exterior error is expressed through two functions of period $2\pi/\Delta$ in $v$. We bounded it using $10^6$ points of one period at which every angle is an exact rational multiple of $\pi$, with every alias of each point bounded. We bounded the norm using $10^5$ gap points and a Bernstein bound. At $\kappa=10^4$ and $\varepsilon=10^{-6}$ we used the function constructed for $\eta=10^{-10}$. We chose the number of terms by the truncation rule of Sec.~S7, minimized over its ellipse parameter. The certified gains are $G\ge1.13497$ at $\kappa=10^4$, $G\ge1.27162$ at $\kappa=10^5$ and $G\ge1.29613$ at $\kappa=10^6$. Sec.~S7 gives the construction, both certificates and their results. The 321 coefficients per point are not printed. They are in the coefficient files on which those certificates operate.

Table~\ref{tab:practical} lists certified target polynomials under A at the three practical points, with error $\varepsilon=10^{-6}$ at $\kappa=10^4$ and $\varepsilon=10^{-10}$ at $\kappa=10^5$ and $10^6$. For each construction it gives the number of odd terms, the certified upper bound on $M/\kappa$ and the certified lower bound on the gain $G$ of \eqref{eq:gain}. For the rescaled minimax baseline, the column $M/\kappa$ gives $\rho_u$ to five decimals. The gain $G$ is calculated using a certified lower bound on $\rho_u$, which Sec.~S7 lists with the other certificates. The CKS rows use the construction underlying the estimates in Section~\ref{sec:subroutine}, namely the polynomial of Childs, Kothari and Somma \cite{childs2017} times a rectangle function. Its normalization $1/(2\kappa)$ gives the entry 2 for $M/\kappa$. The number of odd terms is for the CKS polynomial alone, with $b=\lceil\kappa^2\ln(\kappa/\varepsilon)\rceil$, degree $2d+1$ and $d=\lceil\sqrt{b\ln(4b/\varepsilon)}\rceil$. This count leaves out the degree of the rectangle. The CKS cost and the CKS factors are therefore lower bounds. The CKS factor of a row is the CKS cost divided by the cost of the row, rounded down.

For the baseline from prior work \cite{sunderhauf2025}, the CKS factors are 2.27, 1.80 and 1.76 at the three points. For our constructions, the corresponding factors equal these values times $G$. Against the CKS construction the total factors are therefore at least 2.58, 2.29 and 2.29. Our constructions account for 13, 29 and 31 percent of the logarithm of the total improvement factor over CKS.

\subsection{A two-term closed-form family}

The near-optimal constructions need 321 numerically determined coefficients per point. A simpler family uses only the closed form \eqref{eq:closedform}.

\begin{proposition}\label{prop:family}
For $\alpha\ge0$ and integers $n,d\ge1$, the polynomial $q=(1+\alpha)q^*_n-\alpha q^*_{n+d}$ has at most $n+d$ odd terms and satisfies $\|q-1/x\|_{[a,1]}\le(\kappa+1)\beta^n(1+\alpha+\alpha\beta^d)$.
\end{proposition}

\begin{proof}
For $\alpha=0$, $q=q_n^*$ has degree $2n-1$. For $\alpha>0$, its degree is $2(n+d)-1$. On $[a,1]$, $q-1/x=(1+\alpha)(q^*_n-1/x)-\alpha(q^*_{n+d}-1/x)$, so $|q-1/x|\le(1+\alpha)E_n+\alpha E_{n+d}$ by \eqref{eq:closedform}, and $E_n=(\kappa+1)\beta^n$ (Sec.~S1).
\end{proof}

For given $(\alpha,d)$, Proposition~\ref{prop:family} therefore gives an explicit formula for the least $n$ certified to satisfy the error bound. Only $M$ needs numerical evaluation, as the maximum of an explicit function over the gap. Every family optimum reported below is the best value our search found over $(\alpha,d)$ with this $n$. We optimized $(\alpha,d)$ at 47 points, at $\kappa=10^5$ for $L$ from 8 to 40, at $\kappa=10^3$, $10^4$ and $10^6$ for $L\in\{12,20,28,36\}$, and at the practical points $(\kappa,\varepsilon)=(10^5,10^{-10})$ and $(10^6,10^{-10})$. At fixed $\alpha$, the gain rises with $d$ until an abrupt drop occurs when a second local maximum of $|q|$ in the gap exceeds the first. At $\alpha=56$ and $\kappa=10^5$, for example, $G$ falls from $1.209$ at $d/m_u=0.045$ to $1.127$ at $0.047$. The optimum lies at the edge of this drop. We therefore located it by a nested golden-section search. At $\kappa=10^5$ and $\varepsilon=10^{-10}$ it gives $G=1.2101335$.

The optimal gain depends on $L$ alone to the precision we measured. At $L=20$, 28 and 36 it agrees across $\kappa$ from $10^3$ to $10^6$ within $6\times10^{-5}$. The optimal parameters depend on $\kappa$ more strongly. Our search covered $2\le\alpha\le400$ at every $L$. In this range the family first gives a lower cost than the baseline between $L=16$ and $17$. At $L=16$ the best member found in the searched range has $G=0.9984$, below the baseline $\alpha=0$. The optimal gain rises to $1.2424$ at $L=40$ (Table~\ref{tab:rule}). Across all 35 points where the family costs less than the baseline, the optimal $d$ lies between $0.741\kappa$ and $0.780\kappa$. At $\kappa=10^5$, the optimal $\alpha$ lies within 0.45 percent of its least-squares fit, the line $1.4767\,L+6.463$. At the other condition numbers it lies within 3.6 percent of this line. This gives the rule
\begin{equation}\label{eq:rule}
d=\mathrm{round}(0.76\,\kappa),\qquad \alpha=0.98\,(1.4767\,L+6.463),\qquad
n=\min\bigl\{n:(\kappa+1)\beta^n(1+\alpha+\alpha\beta^d)\le\varepsilon\bigr\},
\end{equation}
for $L\ge17$, with $\alpha$ rounded to six decimals. The factor $0.98$ gives a 2 percent margin before the drop. The factor also leaves a margin of at least 2.6 percent in $d$. Over the 35 points the rule loses at most 0.22 percent of $G$ against the best value found at each point. Outside the fitting set, at $\kappa=10^2$, $10^4$ and $10^7$ and at half-integer $L$ from 17.5 to 39.5, the rule stays within 0.22 percent of the best-found values interpolated in $L$ between the fitting points. A parameter past the drop cannot make the polynomial infeasible, because $n$ and $M$ are computed for the chosen parameters. It only lowers $G$. We tested margins around the rule with $d$ from $0.74\kappa$ to $0.78\kappa$ and $\alpha$ from 0.95 to 1.00 times the fitted line. Across these parameter ranges, the loss is at most 4.23 percent. Parameters further past the drop lose more, as the example at $\alpha=56$ shows.

Table~\ref{tab:rule} lists the best found values and the rule at $\kappa=10^5$. The optimizer evaluates the gap maximum on a uniform grid of 8001 points with golden-section refinement, which gives the columns $\alpha$, $d/\kappa$ and $n/m_u$. The column ``$G$, best found'' is re-evaluated on 400001 gap points (Sec.~S11). The column ``$G$, rule'' is re-evaluated on 40001 points. All values in the table are measured. The two $G$ columns are rounded down. The parameter columns are rounded to the digits shown.

\paragraph{Procedure.} For given $\kappa$ and $\varepsilon$ with $L\ge17$ the polynomial and its normalization follow in five steps.
\begin{enumerate}
\item Allocate the polynomial error $\varepsilon_{\rm poly}$ and use it as $\varepsilon$ in the rule. Put $a=1/\kappa$ and $\beta=(1-a)/(1+a)$, and take $d$, $\alpha$ and $n$ from \eqref{eq:rule}.
\item Form $q=(1+\alpha)q^*_n-\alpha q^*_{n+d}$ from \eqref{eq:closedform}. Proposition~\ref{prop:family} gives $|q-1/x|\le\varepsilon$ on $[a,1]$, so the error needs no check.
\item Compute $M=\max_{[0,1]}|q|$. For the rule polynomials we certified, the maximum lies inside the gap near $x=a/2$, where \eqref{eq:closedform} is accurate in double precision. Lemma~\ref{lem:cert} gives a rigorous bound from values evaluated in ball arithmetic on a fine grid.
\item Apply $c\,q$ with $c=1/(M(1+\xi))$. The margin $\xi$, for example $10^{-3}$, leaves the target below the norm bound one and divides $G$ by $1+\xi$. It is a normalization margin, not an error reserve. In unscaled reciprocal-error units, allocate $\varepsilon_{\rm poly}+\varepsilon_{\rm phase}+\varepsilon_{\rm synthesis}\le\varepsilon_{\rm total}$. The last two terms include the factor $1/c$ that converts implementation errors to these units.
\item For Chebyshev coefficients or pointwise values, use the evaluation forms of Sec.~S1. Proposition~S1.3 applies inside the gap. Outside it, the stated error bound controls the subtraction from one. The form \eqref{eq:closedform} loses accuracy near $x=0$.
\end{enumerate}
Reserving part of $\varepsilon$ for the phase factors and for gate synthesis raises the cost of the rule polynomial and of the rescaled baseline. Over the range of Table~\ref{tab:rule} it also raises their ratio, the gain. Running \eqref{eq:rule} with half of $\varepsilon$ raises $L$ by $\ln2$. By interpolation in that table the gain of the rule then rises by 0.3 to 1.1 percent.

At $\varepsilon=10^{-10}$ the rule gives $\alpha=56.317083$, $d=76000$ and $n=1939103$ at $\kappa=10^5$. At $\kappa=10^6$ it gives $\alpha=59.649306$, $d=760000$ and $n=20570651$. A separate interval computation, run without knowledge of our floating-point values, certified $G\ge1.207619$ and $G\ge1.222085$ (Table~\ref{tab:practical}). It bounds the gap maximum from $10^5$ exact rational points per function in 256-bit ball arithmetic \cite{johansson2017} with the second-derivative bound of Appendix~\ref{app:cert}, at a relative slack of about $2\times10^{-8}$. Our floating-point values agree to seven digits. At $\kappa=10^4$ and $\varepsilon=10^{-6}$ the rule gives $\alpha=39.655969$, $d=7600$ and $n=134623$. We certified this point with a copy of the same code and obtained $G\ge1.105465$. The copy first reproduced the certificate at $\kappa=10^5$ exactly. Sec.~S7 lists the parameters and results of all three certificates. On a logarithmic scale the rule recovers 78.5 and 77.3 percent of the certified gain of the near-optimal constructions at $\kappa=10^5$ and $10^6$. The family optimum recovers 78 and 79 percent of the linear-program optimum at $L=20.7$ and $23.0$.

\begin{table}[t]
\caption{Certified target polynomials under criterion A.}
\label{tab:practical}
\centering
\begin{tabular}{llcccc}
\toprule
$\kappa$ & construction & odd terms & $M/\kappa$, upper & $G$, lower & CKS factor \\
\midrule
$10^4$ & CKS times rectangle, $1/(2\kappa)$ & $290\,933$ & 2 & -- & 1 \\
 & rescaled minimax (baseline) & $m_u=115\,130$ & $\rho_u=2.21947$ & 1 & 2.27 \\
 & near-optimal, 321 coefficients & $139\,671$ & 1.611930 & 1.13497 & 2.58 \\
 & two-term family, rule \eqref{eq:rule} & $142\,223$ & 1.625258 & 1.10546 & 2.51 \\
\midrule
$10^5$ & CKS times rectangle, $1/(2\kappa)$ & $4\,196\,178$ & 2 & -- & 1 \\
 & rescaled minimax (baseline) & $m_u=1\,726\,940$ & $\rho_u=2.69590$ & 1 & 1.80 \\
 & near-optimal, 321 coefficients & $2\,224\,248$ & 1.646030 & 1.27162 & 2.29 \\
 & two-term family, rule \eqref{eq:rule} & $2\,015\,103$ & 1.913170 & 1.20761 & 2.17 \\
\midrule
$10^6$ & CKS times rectangle, $1/(2\kappa)$ & $45\,279\,290$ & 2 & -- & 1 \\
 & rescaled minimax (baseline) & $m_u=18\,420\,682$ & $\rho_u=2.78144$ & 1 & 1.76 \\
 & near-optimal, 321 coefficients & $23\,437\,824$ & 1.686593 & 1.29613 & 2.29 \\
 & two-term family, rule \eqref{eq:rule} & $21\,330\,651$ & 1.965489 & 1.22208 & 2.16 \\
\bottomrule
\end{tabular}
\end{table}

\begin{table}[t]
\caption{The two-term family at $\kappa=10^5$, best found and rule.}
\label{tab:rule}
\centering
\begin{tabular}{cccccc}
\toprule
$L$ & \shortstack{$\alpha$\\best found} & \shortstack{$d/\kappa$\\best found} & \shortstack{$n/m_u$\\best found} & \shortstack{$G$\\best found} & $G$, rule \\
\midrule
17 & 31.49 & 0.768 & 1.216 & 1.0175 & 1.0165 \\
20 & 36.09 & 0.764 & 1.190 & 1.0669 & 1.0654 \\
24 & 42.03 & 0.764 & 1.165 & 1.1186 & 1.1167 \\
28 & 47.81 & 0.766 & 1.146 & 1.1593 & 1.1571 \\
32 & 53.87 & 0.763 & 1.131 & 1.1922 & 1.1898 \\
36 & 59.44 & 0.767 & 1.119 & 1.2195 & 1.2169 \\
40 & 65.58 & 0.764 & 1.110 & 1.2424 & 1.2398 \\
\bottomrule
\end{tabular}
\end{table}

\subsection{Phase factors}\label{sec:phases}

QSVT applies a polynomial through a sequence of $N+1$ phase angles, where $N$ is the degree. Computing these angles is a separate numerical problem, which becomes harder with the degree and as $\max|p|$ approaches one. We computed them for the rule polynomials \eqref{eq:rule} at $L=20$ and $\kappa\in\{100,500,2000\}$, scaled to $p=q/(M(1+\xi))$ with a margin $\xi$. We used Newton's method for symmetric quantum signal processing \cite{dong2023newton} in the convention of \cite{dong2021qsp}, with an exact Jacobian, a dense linear solve on 8 threads, initialization at $(\pi/4,0,\dots,0)$ and no continuation. Table~\ref{tab:phases} lists the results. The error column gives the largest difference between the polynomial realized by the angles and the target $q/(M(1+\xi))$ over the verification points. We evaluated this difference in double precision. The last column is the largest value of $|M(1+\xi)p_{\rm realized}-1/x|/\varepsilon$ over the verification points in $[a,1]$.

The realized polynomial was evaluated from the angles at $4N+2$ Chebyshev-angle points, 12000 points in the gap and 1201 points clustered at $x=a$. Separately, we evaluated it from the phase files with our own code, against a target computed with the cancellation-free form of Sec.~S1. The differences were at most $4.3\times10^{-12}$. A 50-digit evaluation at 200 points of each run gives differences to the target of at most $2.5\times10^{-12}$. In every case $|p|\le1$ at all verification points. The rescaled polynomial $M(1+\xi)p$ meets the error $\varepsilon$ at the verification points in $[a,1]$. The margin divides $G$ by $1+\xi$. Sec.~S11 gives the details.

\begin{table}[b]
\caption{Phase factors of the rule polynomials at $L=20$.}
\label{tab:phases}
\centering
\begin{tabular}{cccccc}
\toprule
$\kappa$ & degree $N$ & $\xi$ & Newton iterations, time & error & realized error$/\varepsilon$ \\
\midrule
100 & 2531 & $10^{-3}$ & 9, 3.9 s & $3.4\times10^{-13}$ & 0.99312 \\
100 & 2531 & $10^{-4}$ & 11, 4.6 s & $3.4\times10^{-13}$ & 0.99311 \\
500 & 12653 & $10^{-3}$ & 10, 116 s & $1.7\times10^{-12}$ & 0.99794 \\
500 & 12653 & $10^{-4}$ & 11, 128 s & $1.7\times10^{-12}$ & 0.99794 \\
2000 & 50609 & $10^{-3}$ & 10, 2298 s & $5.7\times10^{-12}$ & 0.99952 \\
\bottomrule
\end{tabular}
\end{table}

Phase factors computed for an inversion polynomial itself, rather than estimated by extrapolation, have been obtained up to $\kappa=650$ with the normalization fixed at $0.125/\kappa$ \cite{novikau2024}. The same work extrapolates beyond this condition number. Ni and Ying demonstrated a fast phase-finding method on other polynomials at degrees of about $10^5$ \cite{ni2024fast}. Berntson and S\"underhauf computed complementary polynomials, the first step of such methods, up to degree $10^7$ \cite{berntson2024complementary}. Lapworth and S\"underhauf computed 14011 phase factors for an inversion polynomial at an effective condition number of 2500 and error $10^{-2}$ with the Remez method of QSPPACK \cite{lapworth2025}. They used them in an emulated solve. The Newton iteration used here solves a dense system of order about $N/2$ in each step. Each step took about 230 s at $N=50609$. The degrees of the certified points, $2.8\times10^5$ to $4.7\times10^7$, therefore need one of those faster methods.

\paragraph{The certified point at $\kappa=10^4$.} We computed the phase factors of the rule polynomial of Table~\ref{tab:practical} at $\kappa=10^4$ and $\varepsilon=10^{-6}$, of degree 284445, scaled by its certified $M$ with $\xi=10^{-3}$. We used the Weiss algorithm followed by the inverse nonlinear fast Fourier transform \cite{ni2024fast,ni2025inverse}. We ran the code of Ni and Ying on 8 threads, which took 30 s. We checked the result with our own code. At this degree a double-precision evaluation of the realized polynomial itself erred by up to $1.7\times10^{-11}$ at the points where we compared it with 50-digit values. The checks below therefore use 50 digits or 128 bits. At the points evaluated in this way, the realized polynomial differs from the target by at most $6.3\times10^{-12}$ in the gap and by at most $1.8\times10^{-13}$ on $[a,1]$. A scan at 20 points per oscillation found 15221 local maxima of the target error ratio $|q-1/x|/\varepsilon$ on $[a,1]$ above 0.99. We refined these locations and evaluated the realized polynomial there. The largest realized error ratio $|M(1+\xi)p-1/x|/\varepsilon$ at these points is below 0.999866. The largest target error ratio at these points is below 0.999868. The realized polynomial therefore meets the error $\varepsilon$ at these points. At every point evaluated, $|p|<0.99901$. These evaluations use high-precision arithmetic at selected points. They are not interval enclosures, because enclosures widen beyond use over a product of this length. This run allocated $\varepsilon_{\rm poly}=\varepsilon$ with no phase reserve. The realized-error evidence is the direct check at the refined points. By the triangle inequality, the sampled discrepancy $1.8\times10^{-13}$ corresponds to about $0.0029\varepsilon$ after multiplication by $M(1+\xi)$, about 22 times the unused target error $1.33\times10^{-4}\varepsilon$. A design with $\varepsilon_{\rm poly}=\varepsilon/2$ would leave a reserve about 170 times the sampled discrepancy after conversion to reciprocal-error units at the same phase accuracy. This comparison does not certify an error between sampled points. Sec.~S11 gives the details.

\begingroup
\widowpenalty=\csname @M\endcsname\relax
Under criterion A, at $\kappa=10^5$ and $10^6$ with $\varepsilon=10^{-10}$, the polynomial divided by its maximum $M$ needs to be realized to an absolute error below $\eta$, that is below $10^{-15}$ and $10^{-16}$, because $M$ exceeds $\kappa$. These thresholds are below the double-precision rounding error accumulated over the $3.4\times10^6$ to $4.7\times10^7$ factors of these polynomials. We therefore expect phase finding there to need extended-precision arithmetic, for the rescaled minimax polynomial as well. We have not computed phase factors at these two points. The baseline needs phase finding at comparable degrees, 230259 at $\kappa=10^4$ and $3.7\times10^7$ at $\kappa=10^6$. The certified constructions have 1.15 to 1.29 times its degree.

\par\endgroup
\subsection{Finite construction under residual error}\label{sec:residual_construction}

For criterion B we solve a finite problem directly at the target condition number $K$. With $a=1/K$, even $\varpi$ and $m=J\varpi$, put
\begin{equation*}
w(t)=\sum_{j=0}^{J}c_jT_{j\varpi}(t),\qquad
w_{\xi_{\rm B}}(t)=(1-\xi_{\rm B})w(t)+\xi_{\rm B}\frac{T_{J\varpi}(t)}{T_{J\varpi}(t(0))},\qquad
\xi_{\rm B}=\frac{3}{1000000},
\end{equation*}
and $q(x)=(1-w_{\xi_{\rm B}}(t(x^2)))/x$. The last coefficient of $w$ enforces $w(t(0))=1$ exactly, so $q$ is an odd polynomial. On $t=\cos\theta$ the fit is a cosine polynomial of degree $J$ in $\varpi\theta$. In the gap it is $\sum_j c_j\cosh(ju)$, where $u=\varpi\operatorname{arcosh}(-t)$. We optimize the coefficients at finite $K$ with $J=160$. No approximation based on a scaling limit or truncation is used. The common repair weight above is part of the definition of every published polynomial.

Two separately written Arb programs certify all six published instances at that weight. One bounds the fit on a uniform phase grid, and the other uses adaptive Taylor intervals and derivative signs. Each bounds the gap independently by Lemma~\ref{lem:cert}. Table~\ref{tab:residual_cert} prints the weaker bounds, rounded outward. These target polynomials have degree $2m-1$. The gain is relative to the known residual-minimax polynomial at $m_R$. The error targets differ from those of Table~\ref{tab:practical}. No phase factors are claimed for the polynomials in Table~\ref{tab:residual_cert}. Sec.~S12 defines the coefficients and gives the fit repair, both certificates and the three benchmark instances. The coefficient values are in the coefficient file \texttt{anc/verification\_set/\allowbreak criterion\_b/\allowbreak coefficients.json}. For normalized-state error in the 2-norm, the savings are at least 8.7 percent at $K=10^4$ with error at most $10^{-8}$, and at least 13.1 percent at $K=10^5,10^6$ with error at most $10^{-10}$. The first gain lies between the small-$\kappa$ LP values at $\delta=10^{-8}$ and $3\times10^{-9}$. This is consistent with a scaling picture but does not prove one.

\begin{table}[t]
\caption{Certified target polynomials under criterion B.}
\label{tab:residual_cert}
\centering
\begin{tabular}{lllll}
\toprule
$K$ & $\delta$ & $m$ & $M/K$, upper & $G_B$, lower \\
\midrule
$10^4$ & $5\times10^{-9}$ & 122560 & 1.505454 & 1.095883 \\
$10^5$ & $5\times10^{-11}$ & 1464640 & 1.635012 & 1.151636 \\
$10^6$ & $5\times10^{-11}$ & 14635200 & 1.636157 & 1.151710 \\
\bottomrule
\end{tabular}
\end{table}

\section{Windowing}\label{sec:windowing}

S\"underhauf et al.\ describe multiplication by a window polynomial as an alternative to rescaling \cite{sunderhauf2025}. In this section criterion A applies. A window multiplies a minimax polynomial that already meets the error. Hence $n\ge m_u$ throughout. For a fixed $q^*_n$ and an even window $W$ with $k+1$ Chebyshev terms, the product $q^*_nW$ is linear in $W$. We can therefore find the best window for given $(n,k)$ by a linear program. Apart from evenness, the window has no structural restriction beyond its number of terms. At $\kappa=20$ and $\eta\in\{10^{-8},10^{-9},10^{-10}\}$, over $n/m_u$ from 1 to 1.1 and $k/m_u$ from 0.02 to 0.3 (120 cases, all feasible and converged), the windowed cost $(n+k)\max|q^*_nW|$ never fell below the rescaled cost. Every cost ratio in this section is the windowed cost divided by the rescaled cost. In these cases it was at least 1.019. A window lowers $M$ appreciably only once $k$ exceeds $m_u$. At $n=m_u$, $\eta=10^{-8}$ and $k/m_u=1.5$, $M/\kappa$ falls from 2.0 to 1.30. The cost ratio is then 1.63. The reason is that the window needs to stay within about $\varepsilon x$ of one at each $x$ of $[a,1]$ while dropping inside the gap. Near the gap edge this tolerance is about $\eta$. The required window is therefore a high-accuracy step approximation with a degree of the same order as the polynomial approximating the inverse.

The absence of savings in the tested cases does not establish that windowing never lowers the cost. When $\varepsilon$ exceeds $E_{m_u}$, a constant window below one already lowers $M$ at no degree cost. Even without that slack, there is a counterexample at $\kappa=2$ and $\varepsilon=3^{1-65536}$, with cost ratio below 0.76. As $\eta\to0$, the best windowed cost over the rescaled cost tends to zero like $L^{-1/2}$. In the iterated limit, $\kappa\to\infty$ first and $\eta\to0$ afterwards, $\sqrt{L}$ times this ratio lies between $6.5355$ and $10.2601$. The lower coefficient is that of the unrestricted bounded optimum, because a windowed product is itself a feasible polynomial. For the specific Chebyshev window family $xq_{R,\ell}(x)=1-T_\ell(t(x^2))/T_\ell(t(0))$, the cost tends to $1.673$ times the rescaled cost in the same iterated limit. These asymptotic limits give no bound at finite $L$, so the observed absence of savings at the tested accuracies rests on the 120 cases above. At the points $\kappa=10^4$ to $10^6$ of Table~\ref{tab:practical} we have no measurement of the best window. At $\kappa=10^5$ and $10^6$, the cost of every member of the Chebyshev window family above exceeds 1.70 times the cost of the rescaled minimax polynomial. This lower bound applies only to that family, not to the best window over all allowed windows. In the iterated limit stated above, the best window costs at most $10.2601/6.5355\le1.570$ times the bounded optimum. Sec.~S8 gives the proofs and the measurements.

\section{Where the normalization matters}\label{sec:subroutine}

For a normalized solve with amplitude amplification, the relevant gain is $G_B$ of Section~\ref{sec:residual_construction}. The criterion-A gain \eqref{eq:gain} applies where absolute inverse operator accuracy is required. When a block-encoding of $A^{-1}$ or of a resolvent is a factor inside a larger algorithm, $1/c$ enters the total cost differently. Table~\ref{tab:uses} separates these uses at $\kappa=10^6$. The two solve rows use criterion B and compare the sparse construction with $q_{R,m_R}$ at $\delta=5\times10^{-11}$. Their normalized-state 2-norm error is at most $10^{-10}$. The other rows use criterion A and compare the near-optimal construction and the two-term rule with the rescaled minimax polynomial. Under A the absolute operator-error bound is $\varepsilon=10^{-10}$ per inverse factor. Where two saving factors are listed, the first corresponds to the near-optimal construction and the second to the two-term rule. The numerical saving factors are certified lower bounds in this model from Tables~\ref{tab:practical} and~\ref{tab:residual_cert}. Relative to the rescaled minimax polynomial under criterion A, the near-optimal construction raises $c$ by the factor $\rho_u\kappa/M=1.649$ at $1.2724$ times the degree. A cost proportional to $c^{-p}$ therefore falls by the factor $1.649^p/1.2724$.

Without amplitude amplification a solve succeeds with probability proportional to $c^2$. The expected number of repetitions and the cost then scale as $c^{-2}$. This route matters when amplification is unavailable, although it costs more than the amplified approach for a single solve. For the unamplified solve under criterion B, the saving factor is $G_B^2m/m_R$. For the $K=10^6$ polynomial of Table~\ref{tab:residual_cert}, this factor is at least $1.59$ against $q_R$ at $m_R$ of Section~\ref{sec:setting}. A product of two inverse block-encodings has the product of their normalizations. Its cost scales as $c^{-2}$ under the amplified-product model used here. If $\|Q_i-A_i^{-1}\|\le\varepsilon$ and $\|A_i^{-1}\|\le\kappa$, the product error is at most $2\kappa\varepsilon+\varepsilon^2$, so the per-factor error is not the product error. At $\kappa=10^6$ and $\varepsilon=10^{-10}$, this bound is approximately $2\times10^{-4}$. A chain of $k$ non-unitary steps carries the $k$-th power of the normalization. Amplifying each step does not remove this power in general, because every round of amplification reflects about the input of that step, which the earlier steps need to prepare again. The criterion-A construction was optimized for a cost linear in $1/c$. For the amplified product, a saving factor of $1.649^2/1.2724$, at least 2.13, is achievable with the construction under criterion A. The best saving factor for this cost model is at least as large. The two-term rule gives $1.22$ for the linear and $1.72$ for the quadratic uses under A. The factors are our arithmetic, not values reported by the cited works.

\begin{table}[t]
\caption{Saving factors in the stated cost model.}
\label{tab:uses}
\centering
\begin{tabularx}{\textwidth}{>{\raggedright\arraybackslash}X>{\raggedright\arraybackslash}Xl}
\toprule
use & dependence on $1/c$ & saving factor \\
\midrule
single solve with amplitude amplification & linear & 1.15 \\
observable estimation, e.g.\ Green's functions \cite{tong2021,nelson2026} & linear, through amplitude estimation & 1.29 / 1.22 \\
solve without amplitude amplification (repeat until success) & quadratic & 1.59 \\
amplified product of two inverse block-encodings & quadratic & 2.13 / 1.72 \\
chain of $k$ non-unitary steps & power that grows with $k$ & grows with the chain \\
\bottomrule
\end{tabularx}
\end{table}

Uniform singular value amplification can increase the normalization after the polynomial has been chosen. It costs $O((\gamma/\theta)\log(\gamma/\varepsilon))$ queries for an amplification factor $\gamma$ and a margin $\theta$ below the norm \cite{gilyen2019}. For a product of two block-encodings, Lapworth and S\"underhauf give conditions under which amplifying the factors before the multiplication is cheaper \cite{lapworth2025}. For two equal factors these conditions reduce to $\gamma>(3/\theta)\log(\gamma/\varepsilon)$. At $\gamma\approx1.6$ and $\varepsilon=10^{-10}$ this condition fails for every $\theta\le1$. Within the cited amplification construction and its approximate cost model, choosing a polynomial with better normalization is therefore cheaper than applying uniform singular value amplification at these parameters.

\paragraph{A scenario based on one published estimate.} Nelson and Baczewski estimate the T-gate cost of Green's functions at 1 percent accuracy \cite{nelson2026}. Their inverse is the Childs--Kothari--Somma polynomial times a rectangle function, with normalization $1/(2\kappa)$. In their cost model the cost per evaluation is proportional to degree divided by normalization, as in \eqref{eq:cost}. The text of that work states the degree as $d+n$. Its Eq.~(D3) runs to $T_{2d+1}$. We use this second reading unless we state otherwise. We compared the costs per evaluation at $\kappa\in\{10,100,1000\}$. We left out the rectangle's degree, so the baseline is a lower bound on their cost. For the 1 percent accuracy target, we assume a polynomial-error budget between $10^{-2}$ and $10^{-4}$ of the largest value $\kappa$ of $1/x$, that is $\eta$ from $10^{-2}$ to $10^{-4}$. Their work does not state the polynomial error it assumes.

The gain $G$ comes from our linear programs, which reach $\kappa=40$. At $\eta$ from $10^{-2}$ to $10^{-6}$, $G-1$ falls by a factor of 5.5 to 12 from $\kappa=10$ to $40$. We therefore use the gains at $\kappa=40$ for all three condition numbers of the scenario. At $\kappa=100$ and $1000$ they may overstate the gain. The assumed gains are $G=1.0016$, $1.0012$ and $1.0008$ at $\eta=10^{-2}$, $10^{-3}$ and $10^{-4}$. The value at $10^{-3}$ is the mean of the values at $10^{-2}$ and $10^{-4}$. At $\eta=10^{-8}$ and $10^{-10}$ the gains are those of Table~\ref{tab:measured}. The log \texttt{recost\_nb.log} lists the gains at $\kappa=10$, $20$ and $40$ with the factors and shares of both readings. 

Replacing their inverse by the rescaled minimax polynomial lowers the cost by a factor of 3.5 to 6.4 at 1 percent accuracy. This factor comes from the minimax polynomial, which is prior work \cite{sunderhauf2025}. Under the assumed gains above, the further cost reduction is $1-1/G=0.08$ to $0.16$ percent. The estimated total saving factor is the saving factor from the rescaled minimax polynomial multiplied by $G$. We measure the contribution of the assumed gain as its share of the logarithm of this total, that is $\ln G$ divided by the logarithm of the total. This share is 0.06 to 0.10 percent at 1 percent accuracy, 6.8 to 7.4 percent at $\eta=10^{-8}$ and 14.0 to 15.2 percent at $\eta=10^{-10}$. Under the $d+n$ reading, the cost-reduction factors from the rescaled minimax polynomial are halved. The share of the assumed gain then rises to 0.11 to 0.17 percent at 1 percent accuracy, 22 to 31 percent at $\eta=10^{-8}$ and 57 to 84 percent at $\eta=10^{-10}$.

\paragraph{Matched evaluations of a published query model.} Lefterovici et al.\ report QSVT query counts for Poisson matrices at normalized-state 2-norm error $10^{-8}$ \cite{lefterovici2026}. Their implementation assumes unit solution norm. It applies its inverse polynomial to $A/d$, so its effective condition number is $K=d\kappa$. Their Fig.~3 plots $d-1$, whereas their query-count calculation uses $d$. Every matched evaluation uses the stored sparsity.

Our independent implementation reproduces 77 of the 90 stored Poisson counts to $10^{-9}$ relative. Two rows at $\kappa=1$ differ by $9.4\times10^{-4}$. For the last 11 rows, the implemented model gives counts 9.4 to about 14600 times the stored values. We do not infer the cause. Exactly two cells with $K\ge10^4$ have all their instances reproduced. They are $80\le\kappa<100$, $d=125$, with $K=11264.275$ and $11345.4625$, and $1500\le\kappa<1520$, $d=7$, with $K=10632.44$.

At $\delta=5\times10^{-9}$, replacing only the inverse polynomial by $q_R$ from prior work gives cost reduction factors of 71.67 to 71.69 and 78.37 in evaluations of the published model as implemented. Using the common-weight polynomials of Section~\ref{sec:residual_construction} instead gives further factors of 1.146 to 1.147 and 1.101, for totals of about 82.2 and 86.3. The common-weight polynomials have certified lower bounds on $G_B$ of 1.0960 to 1.0964 (Sec.~S12). The model saving factors differ from the gain based on degree and normalization because the implemented amplification formula is not proportional to the inverse square root of the success probability. The reported cost-reduction factors are model evaluations, not measured circuit costs or certified end-to-end costs. Appendix~\ref{app:repro} specifies the source and summation conventions, and Sec.~S12 lists the counts.

\section{Limitations}\label{sec:limitations}

For a solve with amplitude amplification, the cost model takes the query cost to be proportional to degree divided by normalization. It does not include the classical computation of phase factors. We computed phase factors only for the two-term family, up to $\kappa=10^4$ (Section~\ref{sec:phases}). For a single linear solve, solvers whose query count is linear in $\kappa$ are known \cite{costa2023,dalzell2024}. The gain therefore matters most where the inverse block-encoding is a subroutine.

The criterion-A linear programs stop at $\eta=10^{-10}$ and $\kappa=40$ because of double precision (Section~\ref{sec:measured}). Between $L=23$ and $34.5$ our finite bounds on the least cost come only from constructions, which give lower bounds on $G$. The certified gains at the practical points are lower bounds on the optimal finite-condition-number gain, whereas the upper bounds of Section~\ref{sec:practicallimit} apply only to the limiting gain. The rule \eqref{eq:rule} was measured at selected points with $L$ from 17 to 40 and $\kappa$ from $10^2$ to $10^7$, and certified at three points. On our integer grid up to $L=16$, no member with $2\le\alpha\le400$ costs less than the plain minimax polynomial. In the reported search, the crossover lies between $L=16$ and $17$.

The upper bound of Theorem~\ref{thm:frontier} rests on a theorem of semiclassical functional calculus \cite[arXiv v2, Theorem~4.5]{kuster2015}, which Appendix~\ref{app:upper} states. The upper-bound construction does not quantify how large $L$ needs to be before the asymptotic estimate is accurate (Section~\ref{sec:theory}). The sign of the finite-$\kappa$ correction at small $\eta$ is open (Section~\ref{sec:practicallimit}). The criterion-B theory remains open, and Table~\ref{tab:residual_lp} contains LP estimates. The saving factors of Table~\ref{tab:uses} come from our cost model. The matched model evaluations cover two clean cells of one benchmark, whose other rows include 13 discrepancies between the stored counts and the implemented model.

\section{Conclusion}

Choosing the degree and the normalization of a QSVT inverse together, instead of rescaling the least-degree minimax polynomial for the same error criterion, lowers the amplified query cost by an amount that depends on the accuracy and on the error criterion. Under absolute reciprocal error, an entire-function extremal problem gives the optimal gain in the limit of growing condition number at fixed scaled error (Section~\ref{sec:theory}). At high accuracy this gain has the leading term $0.1530\sqrt{\ln(\kappa/\varepsilon)}$, which exceeds one only beyond $\ln(\kappa/\varepsilon)=42.7$. At practical accuracies we therefore use certified two-sided bounds to characterize the limiting gain rather than relying on this leading term. By contrast, the certificates of Section~\ref{sec:constructions} give lower bounds on the gain at finite condition numbers. At condition numbers $10^5$ and $10^6$ with absolute reciprocal error $10^{-10}$, the near-optimal constructions save at least 21 percent of the query cost (Table~\ref{tab:practical}). Under residual error, the sparse Chebyshev construction saves at least 13 percent at the same condition numbers with normalized-state error at most $10^{-10}$ (Table~\ref{tab:residual_cert}).

\begingroup
\widowpenalty=\csname @M\endcsname\relax
The error criterion is therefore part of every cost estimate. A normalized solve gains the saving under residual error, whereas uses that need inverse-operator accuracy gain the saving under absolute reciprocal error. Under the amplified-product cost model of Section~\ref{sec:subroutine}, the saving factor for a product of two inverses exceeds that for a single inverse, because the product's cost scales with the square of the inverse normalization. Each factor of such a product needs its own error budget. Realizing these target polynomials in a circuit requires computing phase factors and keeping the phase and synthesis errors within their allocated budgets (Section~\ref{sec:phases}). Under criterion A at condition numbers $10^5$ and $10^6$ we expect phase finding to need extended-precision arithmetic. Computing those phase factors is the next step.
\par\endgroup

\begin{samepage}

\section*{Acknowledgments}
The author formulated the research question and directed the research and manuscript preparation. The author used AI tools to assist with literature research, mathematical analysis, code development, and manuscript preparation. The author takes full responsibility for the content of the paper and the Supplemental Material.

\end{samepage}
\appendix

\section{Proof of Proposition~\ref{prop:sign}}\label{app:sign}

Let $p^\ast$ be the unbounded minimax polynomial. Odd polynomials with $m$ odd terms form a Haar space of dimension $m$ on $[\vartheta,1]$, so $p^\ast-1$ attains $\pm\Lambda$ with alternating signs at $m+1$ points of $[\vartheta,1]$. Every interior alternation point is a local extremum of $p^\ast$ and hence a zero of $(p^\ast)'$. The derivative is an even polynomial of degree $2m-2$, so it has at most $m-1$ positive zeros. If $\vartheta$ or $1$ were not an alternation point, at least $m$ alternation points would be interior, which is too many. Hence both endpoints are alternation points. The $m-1$ interior alternation points are therefore all the positive zeros of $(p^\ast)'$, so $(p^\ast)'$ has no zero in $(0,\vartheta)$. Then $p^\ast$ is monotone on $[0,\vartheta]$ and satisfies $p^\ast(0)=0$, so $|p^\ast|\le1+\Lambda$ on $[0,1]$ and, by oddness, on $[-1,1]$. The polynomial $p^\ast/(1+\Lambda)$ is bounded by one and takes values in $[(1-\Lambda)/(1+\Lambda),1]$ on $[\vartheta,1]$, so its error is $2\Lambda/(1+\Lambda)$.

Conversely, let $p$ be an odd polynomial in the stated class that is bounded by one and has error $E$ on $[\vartheta,1]$. Then $p\in[1-E,1]$ there, and $p/(1-E/2)$ approximates 1 on $[\vartheta,1]$ with error $(E/2)/(1-E/2)$. Hence $\Lambda\le(E/2)/(1-E/2)$, which rearranges to $E\ge2\Lambda/(1+\Lambda)$.

\section{Proof of Proposition~\ref{prop:lower}}\label{app:lower}

Put $h(z)=1-zf(z)$. Since $f$ has type at most $\sigma$ and $|f|\le B$ on $\R$, $\log|h(z)|\le\sigma|\operatorname{Im}z|+\log(1+B|z|)$. Let $\Omega_b=\mathbb{C}\setminus((-\infty,-1]\cup[-b,b]\cup[1,\infty))$. In the upper half-plane define
\begin{equation}
H_b(z)=\operatorname{Re}\Bigl[-i\int_b^z\sqrt{\frac{\zeta^2-b^2}{\zeta^2-1}}\,d\zeta\Bigr],
\end{equation}
with the root tending to one at infinity, and extend $H_b$ evenly to the lower half-plane. On the gaps $b<|x|<1$ the integrand is imaginary, so the normal derivative of $H_b$ vanishes and the even extension is harmonic in $\Omega_b$. Its boundary values are $H_b=0$ on $[-b,b]$ and $H_b=-R(b)$ on $|x|\ge1$. In the gaps it equals $-J_b(s)$ at $b<s<1$. At infinity it satisfies $H_b(z)=|\operatorname{Im}z|-R(b)+O(|z|^{-1})$. Let $g_b(z)=\log|z/b+\sqrt{z^2/b^2-1}|$ be the Green function of $[-b,b]$ with pole at infinity, so $g_b=0$ on $[-b,b]$, $g_b(x)=\arcosh(|x|/b)\ge\log|x|$ for $|x|\ge1$, and $g_b(z)=\log|z|+\log(2/b)+o(1)$. Define $S=\sigma H_b+\log(1+Bb)+g_b$.

On $[-b,b]$, $\log|h|\le\log(1+B|x|)\le S$. On $|x|\ge1$, feasibility gives $|h(x)|=|x|\,|1/x-f(x)|\le|x|e^{-L}$, and $L\ge\sigma R(b)$ gives $\log|h|\le-\sigma R(b)+\log|x|\le S$. At infinity, $S$ matches the linear and logarithmic terms in the growth bound for $\log|h|$, so the subharmonic function $\log|h|-S$ is bounded above in $\Omega_b$. We subtract $\epsilon g_b$ from this difference and apply the maximum principle on truncated domains. We first let the truncation radius tend to infinity with $\epsilon>0$ fixed, and then take $\epsilon\to0$, which gives $\log|h|\le S$ in $\Omega_b$. Evaluating the preceding bound at $z=s$ gives $|1-sf(s)|\le(1+Bb)T$. If $Bs<1$, then $1-Bs\le1-sf(s)\le(1+Bb)T$, which rearranges to \eqref{eq:lower}. If $Bs\ge1$, the right side of \eqref{eq:lower} is at most $1/s\le B$.

\section{Outline of the upper bound in Theorem~\ref{thm:frontier}}\label{app:upper}

Fix $\lambda>1$ and $0<b_-<b_+<b_\lambda$, and put $c=\lambda L/2$. In this appendix $c$ is the bandwidth of the prolate functions, not the normalization of Section~\ref{sec:setting}. The construction gives, for all large $L$, an odd entire function $f_c$ of type at most $\lambda L$ with $|f_c(s)-1/s|\le e^{-L}$ for $s\ge1$ and $\|f_c\|_{\R}\le1/b_-+o(1)$. Letting $b_-\uparrow b_\lambda$ then gives $\limsup_{L\to\infty}\Phi(\lambda L/2,e^{-L})\le1/b_\lambda$, the upper half of Theorem~\ref{thm:frontier}. Sec.~S5.2 gives the proof in five steps.

\emph{Prolate basis} (Sec.~S5.2.1). Let $\psi_{j,c}$ be the regular eigenfunctions, orthonormal on $(-1,1)$, of $\mathcal P_c\psi=-((1-x^2)\psi')'+c^2x^2\psi=\chi_{j,c}\psi$. Each extends to a real entire function of exponential type at most $c$, because it is an eigenfunction of the finite Fourier transform $u\mapsto\int_{-1}^1e^{iczt}u(t)\,dt$, which commutes with $\mathcal P_c$.

\emph{Kernel} (Sec.~S5.2.3). Choose a nonnegative $\varphi\in C_c^\infty((b_-^2,b_+^2))$ with integral one and put $\omega(Z)=\int\varphi(\theta)\mathbf 1_{Z\le\theta}\,d\theta$ for $Z\ge0$. Extend $\omega$ smoothly with compact support to $Z<0$, where the spectrum has no points, and define
\begin{equation}
K_c(x)=\sum_{j\ge0}\omega(\chi_{j,c}/c^2)\,\psi_{j,c}(x)^2,\qquad H_c=K_c/K_c(0),\qquad f_c(z)=\frac{1-H_c(z)}{z}.
\end{equation}
The sum is finite, $K_c(0)>0$ for large $c$, and $H_c$ is even, nonnegative on $\R$ and entire of type at most $2c=\lambda L$ with $H_c(0)=1$. So $f_c$ is odd and entire of the permitted type.

\emph{Interior density} (Sec.~S5.2.2, Lemma~S5.2). For every fixed $r<1$ and integer $\ell\ge0$,
\begin{equation}
K_c(x)=\frac{c}{2\pi}\int_{\R}\omega\bigl((1-x^2)y^2+x^2\bigr)\,dy+O(1)\qquad\text{in }C^\ell([-r,r]).
\end{equation}
The proof expresses $K_c$ as an integral over the rotation orbits of the kernel of $\omega(\mathcal A_\hbar)$ for the operator $\mathcal A_\hbar=-\hbar^2\nabla^2_{\mathbb S^2}+x^2$ on the unit sphere, with $\hbar=1/c$. On rotation-invariant functions $\mathcal A_\hbar$ acts as $c^{-2}\mathcal P_c$. The external input is Theorem~4.5 of K\"uster \cite[arXiv v2]{kuster2015}, which we use with his symbol-class parameter equal to zero, his operator $B$ equal to the identity and his function family equal to $\omega$. For a closed manifold and a smooth real potential $V$, the theorem expresses $\omega(-\hbar^2\nabla^2+V)$ as a sum of localized Weyl quantizations of $\omega$ composed with the classical Hamiltonian $|\zeta|^2+V$, plus terms whose symbols are $O(\hbar)$ with all derivatives and an operator of trace norm $O(\hbar^k)$ for every fixed $k$. The remainder enters only through its operator norm. Step~7 of Lemma~S5.2 combines this norm with a priori Sobolev bounds on the eigenfunctions to obtain a pointwise $C^\ell$ bound on the kernel. The sphere is closed and $x^2$ is smooth on it, so the theorem applies as stated. The leading term gives $H_c\to H_\omega$ in $C^1$ on $[-r,r]$, where $H_\omega$ is a convex mixture of the functions $\sqrt{(\theta-x^2)_+/(\theta(1-x^2))}$ over $\theta\in(b_-^2,b_+^2)$. Hence $|f_c|\le1/b_-+O(1/c)$ on $[0,r]$.

\emph{Forbidden region} (Subsection~S5.2.4 and Lemma~S5.4). Choose $r\in(b_+,1)$. Eigenfunctions with $\chi_{j,c}\le c^2b_+^2$ have no zeros in $(b_+,1)$ and decrease there, so $0\le H_c\le O(1/c)$ on $[r,1]$. An elementary barrier and Sonin estimate (Lemma~S5.4) gives $\sup_{x\ge1}|\psi_{j,c}(x)|\le\Pi_ce^{-c\tau}$ for every fixed $\tau<R(b_+)$, with a prefactor $\Pi_c$ polynomial in $c$. The constants in this estimate are independent of $j$ for all $\chi_{j,c}\le c^2b_+^2$. At most $cb_++1$ modes enter, and $K_c(0)$ is of order $c$. After division by $K_c(0)$, $H_c(x)$ is at most a constant times $(1+c)^2e^{-2c\tau}$ on $[1,\infty)$. Together with the interior and gap bounds, this gives $\|f_c\|_{\R}\le1/b_-+o(1)$.

\emph{Feasibility} (Sec.~S5.2.5). Because $b_+<b_\lambda$, $R(b_+)>1/\lambda$. With $\tau$ strictly between the two, $|f_c(s)-1/s|=H_c(s)/s\le e^{-L}$ for $s\ge1$ and all large $L$.

K\"uster's theorem is the only external theorem in the interior-density step. Step~3 of the proof of Lemma~S5.2 checks its hypotheses. The same proof computes the reduction along the rotation orbits explicitly.

\section{Certifying the gap maximum}\label{app:cert}

\begin{lemma}\label{lem:cert}
Let $q$ be an odd polynomial of degree $N$, $B=a\max_{[-1,1]}|q|$ and $F(s)=a\,q(as)$. Then for $0\le s\le1$,
\begin{equation}
|F''(s)|\le C_DB,\qquad C_D=a^2\Bigl[\frac{N^2}{1-a^2}+\frac{aN}{(1-a^2)^{3/2}}\Bigr].
\end{equation}
If $C_D<8K^2$, the maximum of $|F|$ over a grid of spacing $1/K$ in $[0,1]$ is at most $g$, and $|F|\le1+\eta$ on $[1,\kappa]$, then $B\le\max\{1+\eta,\ g/(1-C_D/(8K^2))\}$.
\end{lemma}

\begin{proof}
Put $x=\sin\theta$ and $u(\theta)=q(\sin\theta)$, a trigonometric polynomial of degree $N$. Bernstein's inequality gives $|u'|\le N\|q\|$ and $|u''|\le N^2\|q\|$, and $q''(x)=[u''(\theta)\cos\theta+u'(\theta)\sin\theta]/\cos^3\theta$, so $|q''(x)|\le[N^2/(1-x^2)+N|x|/(1-x^2)^{3/2}]\,\|q\|$. With $F''(s)=a^3q''(as)$ and $as\le a$ this gives the first bound. Between adjacent grid points, $\pm F$ lies below its linear interpolant plus $C_DB/(8K^2)$, so the maximum of $|F|$ on $[0,1]$ is at most $g+C_DB/(8K^2)$. If $B$ is attained on $[0,1]$ this gives $B\le g+C_DB/(8K^2)$, hence $B\le g/(1-C_D/(8K^2))$ because $C_D<8K^2$, and otherwise $B\le1+\eta$.
\end{proof}

For the rule polynomials at $\kappa=10^5$ and $10^6$, $N$ is about $4\times10^6$ and $4\times10^7$ and $aN$ is about 40. The grid spacing $1/K$ with $K=10^5$ then gives $C_D/(8K^2)\approx2\times10^{-8}$. At $\kappa=10^4$, $N=284445$, $aN\approx28$ and $C_D/(8K^2)\approx1.0\times10^{-8}$. The grid values are evaluated at exact rational points in ball arithmetic.

\section{Reproducibility}\label{app:repro}

The computations used Python 3.12.13, NumPy 2.5.2, SciPy 1.18.0, mpmath 1.3.0, python-flint 0.9.0 and HiGHS 1.15.1, with MATLAB for the large phase calculation. The public set in \texttt{anc/verification\_set/} contains the inputs, programs and expected outputs of every certified result. Its guide is \texttt{anc/verification\_set/README.txt}. Its checksums and licences are in \texttt{anc/verification\_set/SHA256SUMS.txt} and \texttt{anc/verification\_set/LICENSE.txt}. The other records are available from the author on request. They contain the programs, inputs and logs behind measured results. Third-party code and files are excluded. These exclusions cover the phase solver of Ni and Ying and the source files of the matched comparison. The records include the solver's output phases and our checks.

\begin{center}
\small
\begin{tabularx}{\textwidth}{>{\raggedright\arraybackslash}X>{\raggedright\arraybackslash}X}
\toprule
claim & certificate and expected output \\
\midrule
A finite targets, Table~\ref{tab:practical} & \texttt{anc/verification\_set/\allowbreak verify.py a\_targets}, norm upper and gain lower bounds of Sec.~S7.3 \\
limiting frontier and gain & \texttt{anc/verification\_set/\allowbreak verify.py limiting}, the two-sided intervals in the supplement's certificate index \\
elliptic constants & \texttt{anc/verification\_set/\allowbreak verify.py constants\_windows}, the intervals of Corollary~5 \\
B finite targets, Table~\ref{tab:residual_cert} & \texttt{anc/verification\_set/\allowbreak verify.py b\_targets}, fit ratios below one and norm/gain bounds of the six-row certificate table of Sec.~S12 \\
\bottomrule
\end{tabularx}
\end{center}

For the matched model evaluations, we used \url{https://gitlab.ibr.cs.tu-bs.de/alg/qls-comparison}, commit \texttt{e12600d5}. The SHA-256 hashes of \texttt{evaluation/query\_costs.py} and the extracted \texttt{results.txt}, respectively, are
\begin{center}\footnotesize\ttfamily
757b9e4b234090f8b32fa0590484c34a441c0f8b4e7d1b161f131c5c9c5c3eb7\\
2ef7a7f1f11436c8b90f4265a7f09609fd99e8992828a9128fe460796b2adfa7
\end{center}
Our independent model implementation follows the source in skipping the $k=1$ term and dropping the final dyadic block when its sum is below 0.1. The \texttt{README.txt} of the records available on request gives recount commands with these hashed inputs. Their files are not redistributed or executed.

\bibliographystyle{quantum}
\bibliography{refs}

\end{document}